\documentclass[letterpaper,12pt]{article}
\usepackage{setspace}
\usepackage{float}
\usepackage{multirow}
\usepackage{natbib}
\bibpunct{(}{)}{;}{a}{}{,}
\usepackage{graphicx}
\usepackage{amssymb}
\usepackage{amsthm}
\usepackage{multirow}
\usepackage{mathrsfs}
\usepackage{geometry}
\usepackage{amsfonts}

\usepackage{pdflscape}
\usepackage{rotating}
\usepackage{url} 
\usepackage{appendix}

\usepackage{geometry}
\usepackage{tikz}
\usetikzlibrary{positioning}
\usetikzlibrary{arrows}
\usetikzlibrary{automata}
\usepackage{syntonly}

\usepackage{tikz}
\usetikzlibrary{positioning}
\usetikzlibrary{arrows}
\usetikzlibrary{automata}
\usepackage{epstopdf}
\usepackage{adjustbox}
\usepackage{array}

\newcolumntype{R}[2]{%
    >{\adjustbox{angle=#1,lap=\width-(#2)}\bgroup}%
    l%
    <{\egroup}%
}
\newcommand*\rot{\multicolumn{1}{R{45}{1em}}}

\usepackage[usenames,dvipsnames]{pstricks}
\usepackage{epsfig}
\usepackage{pst-grad}
\usepackage{pst-plot}

\usepackage{epstopdf}
\usepackage{amsmath}

\usepackage{threeparttable}
\usepackage{booktabs}
\usepackage{multirow}

\usepackage{lipsum}

\newcommand{\beq}{\begin{equation}}
\newcommand{\eeq}{\end{equation}}

\def\eq#1/{(\ref{e:#1})}
\def\Section#1/{Section~\ref{s:#1}}
\def\Table#1/{Table~\ref{t:#1}}
\def\Figure#1/{Figure~\ref{f:#1}}

\newcommand{\bm}[1]{\mbox{\boldmath$#1$}}

\def\theequation{\arabic{equation}}

\newcommand{\specialcell}[2][c]{%
  \begin{tabular}[#1]{@{}c@{}}#2\end{tabular}}

\newcommand{\specialcella}[2][l]{%
\begin{tabular}[#1]{@{}l@{}}#2\end{tabular}}

\renewcommand{\theequation}{\arabic{equation}}

\newtheorem{assumption}{Assumption}[section]

\input epsf

\newcommand{\blind}{1}

\begin{document}

\def\spacingset#1{\renewcommand{\baselinestretch}%
{#1}\small\normalsize} \spacingset{1}


\if1\blind
{
  \title{\bf Data fusion for predicting long-term program impacts 
  }
  \author{Michael W.~Robbins, Sebastian Bauhoff, Lane Burgette\thanks{
Michael W.~Robbins and Lane Burgette are Senior Statisticians with the RAND Corporation, Pittsburgh, PA 15213 (E-mail:~{\em mrobbins@rand.org}, {\em burgette@rand.org}). Sebastian Bauhoff is Assistant Professor of Global Health and Economics (E-mail:~{\em sbauhoff@hsph.harvard.edu}) with Harvard T.~H.~Chan School of Public Health, Boston, MA 02115.  The authors acknowledge funding from grant R21AG058123 from the National Institutes of Health.}
}
  \maketitle
} \fi

\if0\blind
{
  \bigskip
  \bigskip
  \bigskip
  \begin{center}
    {\LARGE\bf Data fusion for predicting long-term program impacts 
    }
\end{center}
  \medskip
} \fi

\bigskip
\begin{abstract}
Policymakers often require information on programs' long-term impacts that is not available when decisions are made.  
We demonstrate how data fusion methods may be used address the problem of missing final outcomes and predict long-run impacts of interventions before the requisite data are available.  We implement this method by concatenating data on an intervention with auxiliary long-term data and then imputing missing long-term outcomes using short-term surrogate outcomes while approximating uncertainty with replication methods. We use simulations to examine the performance of the methodology and apply the method in a case study. Specifically, we fuse data on the Oregon Health Insurance Experiment with data from the National Longitudinal Mortality Study and estimate that being eligible to apply for subsidized health insurance will lead to a statistically significant improvement in long-term mortality. 
\end{abstract}

\noindent%
\vfill

\newpage
\spacingset{1.5} 

\section {Introduction} \label{sec1}

Information about the long-term effects of policies and interventions is highly valuable but often unavailable when it would be most useful.  One challenge is timing, as decisions about continuing or scaling interventions are made before data on long-term impacts are available \citep{Athey2019}.  Another challenge is that many long-term studies will never be performed so that there is relatively little evidence on the long-term impacts of existing interventions that could inform revisions or the design of future policies.  That is because the costs of follow-up data collection often increase over time while the benefits of evidence may decline as interventions become established or are superseded.

Analysts commonly address this problem missing data on long-term impacts by identifying surrogate markers (i.e., intermediate variables) that are measured in the short term, are likely to be affected by the intervention, and are believed to have a casual relationship with the final outcome of interest.  If an effect of the intervention on the intermediate variables can be established, one may transitively infer that the intervention also affects the long-term outcome.  \cite{prentice89} outlines a set of criteria that may be used to conceptualize whether or not an intermediate variable is valid for such inferences. The primary criterion stated therein stipulates that the final (long-term) outcome is independent on the treatment conditional on an intermediate (short-term) outcome.

%

In order to extend the idea of surrogacy to estimating long-term impacts (instead of simply hypothesizing that such impacts exist), a common workaround is to extrapolate the short-term effects via the ``product of estimates’’ approach.  This involves multiplying regression coefficients describing the intermediate effect with the estimated relation of the intermediate and final outcomes of interest.  For example, \cite{cummings1989cost} multiply the estimated short-term impact of physician counseling on smoking cessation with the estimated increase in life expectancy for individuals who quit smoking.  This technique assumes that the treatment effect is homogeneous and that the two samples are comparable and, as \citet{Athey2019} note, requires surrogate validity in the sense of \cite{prentice89}, which can fail if there is unmeasured confounding between the intermediate and final outcomes.  Importantly, this approach imposes linear models between the sequence of variables, and statistical inference under the approach is often carried out in an {\em ad-hoc} manner, e.g., by multiplying the endpoints of two parameters' 95\% confidence intervals.  \citet{Athey2019} propose an alternative approach that predicts the final outcome based on a ``surrogate index'' consisting of multiple intermediate outcomes that, together, may be more likely meet the surrogacy assumption.  This approach also allows for out-of-sample validation, and dynamic and heterogeneous treatment effects.  


We propose to use data fusion 
to estimate longer-term impacts using information available in the short-run.  Data fusion addresses the problem of missing data on the final outcomes by concatenating data on short-run impacts with auxiliary data on the relation of the intermediary and final outcomes.  Data fusion provides a structured way to match the short and long-run samples, accommodates effect heterogeneity and allows for rigorous inference.  It also exposes the underlying assumptions of such predictions and indicates possible empirical tests.  In addition, data fusion can be implemented with a range of established matching methods and does not impose specific functional forms as the surrogacy index by \cite{Athey2019}.  



Data fusion has a long history in the statistics literature.  Early work of \cite{rubin1986statistical} suggests concatenating the two data files such that data elements that are common to both data sources are observed for all individuals.  Variables that are only observed in one data source are considered missing for individuals in the other data source. Multiple imputations of the missing variables are then produced, and models can be estimated using complete data methods.   Data fusion has been applied in fields ranging from marketing \citep[e.g.,][]{van2008proof} to education \citep[e.g.,][]{kaplan2013data} and remains an area of active methodological research \citep[e.g.,][]{gilula2006direct, reiter2012bayesian, qian2014brand, fosdick16, schifeling19}.  We are not aware of its use in the context of estimating the long-term impacts of policy interventions, as we discuss in this paper.

Herein, we derive the assumptions under which data fusion can be used to estimate long-term impacts. We then outline the proposed methodology, which involves multiply imputing the long-term outcomes after concatenating the two data files. However, we show that known combining rules for multiple imputation fail in this setting and, as such, we discuss the use of the jackknife and bootstrap for approximating uncertainty in point estimators and outline their comparative advantages and disadvantages. Specifically, the jackknife  mandates a larger number of multiple imputations.
This finding offers a new contribution to existing literature on the use of replication methods with imputed data.

To illustrate the utility of data fusion in this context, we consider the long-term effects of health insurance. Insurance can mitigate financial risk and improve access to health care services, which in turn can improve long-term outcomes such as mortality. One important recent study, the Oregon Health Insurance Experiment (OHIE), leveraged a dedicated lottery in 2008 that awarded applicants the opportunity to apply for subsidized health insurance.  The OHIE has yielded rigorous evidence on the short-term effects of having health insurance \cite[e.g.,][]{finkelstein2012oregon, baicker2013oregon}; winning the lottery was shown to be associated with improved short-term health and financial measures. However, while the longer-term impacts of health insurance are likely important \citep{finkelstein2008what}, they are less well understood because the data required for such assessments are generally not available (e.g., the OHIE has not included long-term follow-up).

We apply the proposed data fusion methodology in an effort to quantify the long-term effect of owning health insurance on mortality.  This is accomplished by fusing 
the OHIE with data from a long-term mortality study.  We estimate that winning the lottery is associated with reduced risk of mortality within 11 years.  


\section {Data Fusion for Estimating Long-Run Impacts} \label{sec2}

We begin our exposition on the relevant methodology with definitions and notation.  The intervention influences the {\bf intermediate variables X} that are measured in the short-term and are on the causal pathway to the {\bf final outcomes} ${\bf Y}$, which are the long-run outcomes of interest.  The data required to estimate the long-run impact come from two separate datasets that are incomplete but complementary.  First, the {\bf intervention dataset} contains a binary measure for the individual-level treatment assignment ${\bf Z}_{\rm int}$  and the intermediate variables ${\bf X}_{\rm int}$, but does not contain the final outcomes.  Second, the {\bf outcomes dataset} contains the intermediate variables ${\bf X}_{\rm out}$ and the final outcomes ${\bf Y}_{\rm out}$ for a sample of individuals who were not treated.  By fusing these two datasets at the individual level we can obtain data on the (latent) outcome variable ${\bf Y}_{\rm int}$ that is missing from the intervention dataset.

Note that other work \citep[e.g.,][]{prentice89, wang21} uses terminology such as ``surrogate marker'' in reference to what we refer to as intermediate variables.  We prefer the latter as the former was introduced in the context of using ${\bf X}$ {\em in place of} the final outcome ${\bf Y}$ for analysis.  Here, we use ${\bf X}$ as an intermediate step on the pathway to the final outcome. Furthermore, as we illustrate in Section \ref{examples} it may also be important to include covariates that are not directly impacted by the intervention in the set ${\bf X}$---such variables are not accurately described as surrogates for the final outcome.



\subsection{Assumptions for Data Fusion}

As the data fusion setup mandates that the intervention and final outcomes are not simultaneously observed, inferences regarding the influence of the intervention on the outcomes are subject to assumptions about the relationship between these variables (conditional upon the jointly observed intermediate variables).  Letting $f(\cdot)$ represent general notation for a probability distribution, we assume the following.







\begin{assumption} \label{assump1}
\underline{Surrogate validity}: Conditional upon the intermediate variables, the unobserved outcome measurements for individuals within the intervention dataset are independent of the intervention.  Specifically,
\[
f({\bf Y}_{\rm int}| {\bf X}_{\rm int}, {\bf Z}_{\rm int}) = f({\bf Y}_{\rm int} | {\bf X}_{\rm int}).
\]
\end{assumption}

This assumption aligns with the so-called Prentice criteria for surrogacy \citep{prentice89, berger04} and is akin to ignorability assumptions commonly used in propensity scoring procedures \citep{rosenbaum1983central}, non-response adjustments \citep{little03}, and sample blending \citep{robbins21}---it is also analogous to
the missing at random assumption commonly used when analyzing data containing missingness \cite{little02}. Assumption \ref{assump1} implies that the effect of the intervention on the outcome is explained entirely through the intermediate variables.  That is, conditional on the intermediate variables, there is no residual effect of the intervention on the outcome.  \cite{prentice89} outlines other criteria that effectively assert that the treatment will affect the final outcome; however, we do not impose those here as they are not needed to yield an unbiased estimate of the overall treatment effect.

If this assumption is not satisfied, estimates of the treatment effect that are found using data fusion will likely contain bias \citep[see][who studied a similar setup]{reiter2012bayesian}.
The intervention need not directly impact the outcome for Assumption \ref{assump1} to be violated.  For example, an the presence of an unobserved confounder can violate the assumption, as illustrated in Section \ref{examples}.
As such, concerns about the validity of Assumption \ref{assump1} may be alleviated by including a robust set of intermediate variables and covariates within ${\bf X}$.

We will also need to be able to use the outcomes dataset to gauge an accurate scope of how the intermediate variables and outcome variables relate for individuals within the intervention dataset. This yields the following assumption that establishes that the two datasets are viable for fusion.

\begin{assumption} \label{assump2}

\underline{Data fusion validity}: The conditional distribution of the outcome variables given the intermediate variables for respondents from the outcome dataset is equivalent to its respective distribution for individuals from the intervention dataset.  That is,
\[
f({\bf Y}_{\rm out}| {\bf X}_{\rm out} = {\bf x}) = f({\bf Y}_{\rm int} | {\bf X}_{\rm int} = {\bf x}),
\]
for ${\bf x}$ in the domain of ${\bf X}_{\rm int}$.
\end{assumption}

%
%
Notice that neither of the assumptions impose that the joint distribution of the intermediate variables is the same across the two datasets.  That is, we do not require $f({\bf X}_{\rm out}) = f({\bf X}_{\rm int})$, and consequentially it is not mandated that $f({\bf Y}_{\rm out}) = f({\bf Y}_{\rm int})$.
However, in accordance with the Assumption \ref{assump2}, the domain of $\{{\bf X}_{\rm int},{\bf Y}_{\rm int}\}$ must be a subset of the domain of $\{{\bf X}_{\rm out},{\bf Y}_{\rm out}\}$.

\subsection{Examples to Illustrate the Potential for Bias} \label{examples}

Although the two assumptions are seemingly simple and narrow, a variety of causal pathways that link treatment, intermediate variables, and the outcome can lead to their violation. We illustrate the violation of the assumptions and the resulting bias using a pair of brief examples involving a single outcome variable ($Y$) and a single intermediate variable ($X$) in the presence of a treatment variable ($Z$).  Although we calculate the bias in these simple examples with a product-of-estimates approach, the estimator found using data fusion with imputation (as outlined in Section \ref{impute}) will yield similar bias.
Additional examples are provided in Section \ref{causal} of the Supplemental Materials.

\subsubsection{Example 1}

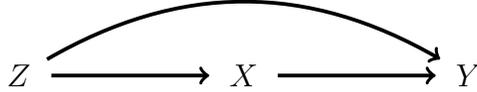
\begin{figure}[!t]
\begin{center}
\begin{tikzpicture}
\matrix[row sep = 0.35cm, column sep = 0.35cm]
{
 \node[circle,thick, draw, white, line width=0.05cm] (n1) {\color{black} $Z$}; & & & & & & \node[circle,thick, draw, white, line width=0.05cm] (n2) {\color{black} $X$}; & & & & & &  \node[circle,thick, draw, white, line width=0.05cm] (n3) {\color{black} $Y$}; \\
};
\draw [->,black, thick, line width=0.05cm] (n1) -- node [above]    {}  (n2);
\draw [->,black, thick, line width=0.05cm] (n2) -- node [above]    {}  (n3);
\draw [->,black, thick, line width=0.05cm] (n1) edge [bend left] node [above]    {}  (n3);
\end{tikzpicture}
\caption{The intermediate variable does not sufficiently model the effect of the treatment on the outcome.} \label{example1}
\end{center}
\end{figure}

To begin, we consider circumstances where the treatment affects the intermediate variable and the outcome is affected by both the treatment and the intermediate variable.  This causal pathway is illustrated graphically in Figure \ref{example1}.
We assume linear conditional relationships between the variables and denote the conditional expectation of the surrogate given the intervention by
\begin{equation} \label{exeq2}
E[X|Z] = \phi_0 + \phi_1 Z.
\end{equation}
Further, the quantity that cannot be estimated directly in the data fusion framework is
\begin{equation} \label{exeq3}
E[Y|Z,X] = \theta_0 + \theta_1 Z + \theta_2 X,
\end{equation}
and Assumption \ref{assump1} is violated whenever $\theta_1 \neq 0$.
However, we are primarily interested in
\begin{equation} \label{exeq1}
E[Y|Z] = \Phi_0 + \Phi_1 Z,
\end{equation}
wherein $\Phi_1$ gives the treatment effect of interest.
In the event that (\ref{exeq2}) and (\ref{exeq3}) are known, it follows that
\begin{equation} \label{exeq4}
\Phi_1 = \theta_1 + \theta_2 \phi_1.
\end{equation}

The parameters $\phi_0$ and $\phi_1$ can be estimated using only the interventions dataset (wherein $X$ is observed but not $Y$).
If $\theta_1 = 0$, Assumption \ref{assump1} holds, and if Assumption \ref{assump2} also holds, we can use the outcomes dataset (wherein $X$ and $Y$ are observed with no treatment) to produce an unbiased estimate of $\theta_2$ which will then yield an unbiased estimate of $\Phi_1$. Alternatively, if $\theta_1 \neq 0$, Assumption \ref{assump1} is violated, and if we then attempt to model $E[Y|Z,X]$ using only the outcomes dataset, we will erroneously suggest that $E[Y|X,Z] = \theta_0 + \theta_2 X$ (assuming the outcomes dataset has essentially been generated via (\ref{exeq2}) and (\ref{exeq3}) with $Z = 0$).  In this case, a product of estimates approach yields
\begin{equation} \label{exeq5}
\widetilde{\Phi}_1 = \theta_2\phi_1
\end{equation}
as the estimated treatment effect. As such, the bias in the product of estimates approach is $\widetilde{\Phi}_1 - {\Phi}_1 = \theta_1$.

\subsubsection{Example 2}

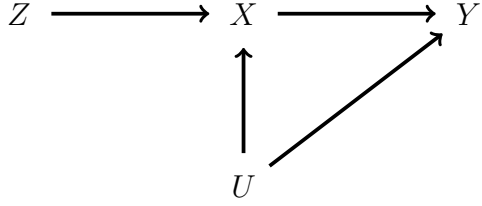
\begin{figure}[!tb]
\begin{center}
\begin{tikzpicture}
\matrix[row sep = 0.35cm, column sep = 0.35cm]
{
 \node[circle,thick, draw, white, line width=0.05cm] (n1) {\color{black} $Z$}; & & & & & & \node[circle,thick, draw, white, line width=0.05cm] (n2) {\color{black} $X$}; & & & & & &  \node[circle,thick, draw, white, line width=0.05cm] (n3) {\color{black} $Y$}; \\
 & & & & & &  & & & & & &   \\
 & & & & & &  & & & & & &   \\
 & & & & & &  & & & & & &   \\
& & & & & & \node[circle,thick, draw, white, line width=0.05cm] (n4) {\color{black} $U$}; & & & & & &  \\
};
\draw [->,black, thick, line width=0.05cm] (n1) -- node [above]    {}  (n2);
\draw [->,black, thick, line width=0.05cm] (n4) -- node [above]    {}  (n2);
\draw [->,black, thick, line width=0.05cm] (n4) -- node [above]    {}  (n3);
\draw [->,black, thick, line width=0.05cm] (n2) -- node [above]    {}  (n3);
\end{tikzpicture}
\caption{Intermediate variable relates to the outcome directly and through an unobserved confounder, and the treatment affects only the intermediate variable.} \label{example2}
\end{center}
\end{figure}

The next example presents circumstances where it is less obvious as to whether or not Assumption \ref{assump1} holds.  Specifically, we extend the aforementioned framework so that the outcome is no longer directly affected by the intervention, but there is an unobserved confounder that affects both the intermediate variable and the outcome.
The resulting data generating mechanism is represented in Figure \ref{example2} and is specifically modeled using
\begin{eqnarray}
\nonumber f(U|Z):& &U = \alpha_0 + \epsilon_u, \\
\nonumber f(X|Z,U): & &X= \beta_0 + \beta_1 U + \beta_2 Z + \epsilon_x, \\
\label{exeq6} f(Y|Z,U,X): & & Y =\gamma_0 + \gamma_1 U + \gamma_2 X + \epsilon_y,
\end{eqnarray}
where $\epsilon_u$, $\epsilon_x$, and $\epsilon_y$ are Gaussian random errors with mean zero and variances $\sigma^2_u$, $\sigma^2_x$, and $\sigma^2_y$, respectively.
We next derive expectations of the form in (\ref{exeq2})-(\ref{exeq1}) under the above system.

First, it is easily established that the model in (\ref{exeq2}) has $\phi_1 = \beta_2$.  Next, using well-known formulas for the conditional distributions of multivariate normal data, we can show that
\begin{equation} \label{exeq7}
E[U|X,Z]=\alpha - c(\beta_0 + \alpha) - \beta_2 c Z + cX,
\end{equation}
where $c = \beta_1 \sigma_u^2/(\beta_1^2 \sigma_u^2 + \sigma_x^2)$.  An implication of (\ref{exeq7}) is that even though the treatment does not directly affect the confounder, a relationship between the two is established through the intermediate variable; therefore, when the confounder is also conditioned on the intermediate variable, a dependence on the treatment manifests.
Plugging (\ref{exeq7}) into (\ref{exeq6}), we see that $E[Y|Z,X]$ as in (\ref{exeq3}) has
\[
\theta_1 = -\gamma_1 \beta_2 c ~~~~\mbox{and}~~~~ \theta_2 = \gamma_1c + \gamma_2.
\]
Since $\theta_1 \neq 0$, Assumption \ref{assump1} is violated and the data fusion estimator will contain bias.

Therefore, the true effect of the treatment on the outcome, as seen in (\ref{exeq4}), is given here by $\Phi_1 = \beta_2 \gamma_2$. If the product-of-estimates expression in (\ref{exeq5}) were used, this treatment effect would be approximated by $\widetilde{\Phi}_1 = \beta_2(\gamma_1c + \gamma_2)$.  As a result, the bias in the data fusion approach is given by
$ 
\gamma_1 \beta_2 {\beta_1 \sigma_u^2}/{(\beta_1^2 \sigma_u^2 + \sigma_x^2)}.
$ 
If all variables are positively correlated (i.e., all coefficients are positive), the bias will be positive.

Note that if either the unobserved confounder does not affect the intermediate variable (i.e., $\beta_1 = 0$) or if it does not affect the outcome (i.e., $\gamma_1 = 0$), then Assumption \ref{assump1} is no longer violated and the bias is zero. In general, a system obeying Figure \ref{example2} will have Assumption \ref{assump1} hold if one of the two arrows stemming from $U$ is disconnected.  

Furthermore, if the confounder is measured and incorporated into the estimation process (i.e., included in ${\bf X}$), Assumption \ref{assump1} will be satisfied and the aforementioned bias removed, as illustrated in Example 6 of the supplemental materials.

\subsection{Imputation} \label{impute}

The above examples illustrate the importance of incorporating covariates into the data fusion process.  However, the rudimentary product-of-estimates approach proves insufficient for situations involving multiple intermediate variables and/or non-linear modeling and does not readily yield statistical inferences. To illustrate the calculation of the treatment effect for general data models, we use a potential outcomes framework.
Let ${X}^{(z)}_i$ and ${Y}^{(z)}_i$ denote the potential outcome for the intermediate and final outcomes under treatment condition ${Z}_i = z$ for case $i$.  Assuming binary treatment (i.e., $z \in \{0,1\}$), the treatment effect is defined as
$ 
\alpha = \mu_1 - \mu_0
$
where $\mu_z=E[{Y}^{(z)}_i] = E[{Y}_i|{Z}_i=z]$.
Letting $f^{(z)}({x})$ denote the density function of ${X}^{(z)}_i$, it follows that
\begin{eqnarray}
\nonumber \alpha & = & \int E[{Y}_i | {Z}_i = 1, {X}_i = {x}]f^{(1)}({x})d{x}
- \int E[{Y}_i | {Z}_i = 0, {X}_i = {x}]f^{(0)}({x})d{x} \\
& = & \int E[{Y}_i | {X}_i = {x}]f^{(1)}({x})d{x}
- \int E[{Y}_i | {X}_i = {x}]f^{(0)}({x})d{x} \label{potential} 
\end{eqnarray}
Note that the second line follows from Assumption \ref{assump1}.  The above quantity is estimable in the context of data fusion since $E[{Y}_i | {X}_i = {x}]$ can be estimated using the outcomes dataset and $f^{(1)}({x})$ and $f^{(0)}({x})$ can be estimated from the interventions data.  We find that imputation through data fusion yields a straightforward approach through which point and interval estimates of $\alpha$ can be extracted for general data classes.


Specifically, 
we suggest imputing the unknown values of ${\bf Y}_{\rm int}$.  The first step in this process is to build a rectangular data file that fuses both the intervention and outcomes dataset and has missing values in place of ${\bf Y}_{\rm int}$. That is, define
\begin{equation} \label{chidat}
\bm{\chi} = \left[
\begin{array}{c}
\bm{\chi}_{\rm out} \\
\bm{\chi}_{\rm int}
\end{array}
\right]
= \left[
\begin{array}{ccc}
\bm{0} & {\bf X}_{\rm out} & {\bf Y}_{\rm out} \\
{\bf Z}_{\rm int} & {\bf X}_{\rm int} & \bm{?}
\end{array}
\right],
\end{equation}
where $\bm{\chi}_{\rm out}$ represents the full outcomes dataset and $\bm{\chi}_{\rm int}$ indicates the full interventions dataset with the values of ${\bf Y}_{\rm int}$ unobserved. Note that cases in the outcomes dataset are effectively untreated (suggesting ${\bf Z}_{\rm out} = \bm{0}$); however, it is possible for the cases in the outcomes data to differ from the control cases from the interventions data all while Assumptions \ref{assump1} and \ref{assump2} hold. The value of ${\bf Z}_{\rm out}$ is not relevant here since at no point will we measure the relationship between ${\bf Z}$ and ${\bf X}$ or ${\bf Y}$ using the outcomes dataset.

We implement this step with the GERBIL algorithm of \cite{robbins20}, as available in the R package \texttt{gerbil} \citep{robbins21} to impute the missing values of ${\bf Y}_{\rm int}$. This procedure uses joint modeling for imputation and is efficiently implemented in data of a general structure and thus avoids the theoretical drawbacks of approaches that use fully conditional specification \citep[FCS,][]{vanbuuren06}, such as \texttt{mice} \citep{vanbuuren10, su11} and IVEware \citep{raghunathan02}.
Further, we investigate multiple imputation \citep{rubin96, little02} as a means of adjusting estimators for imputation error.
%
An advantage of the GERBIL method is that, like techniques employing fully conditional specification such as \texttt{mice}, it enables the user to select the form (in terms of sets of predictors) that is used for each conditional model used for imputation.  

Note, however, that the observed dataset $\bm{\chi}$ contains no information that could be used to establish a direct relationship between ${\bf Z}_{\rm int}$ and ${\bf Y}_{\rm int}$ after conditioning on the intermediate variables. That is, for all cases in $\bm{\chi}$ that have a non-missing value of ${\bf Z}_{\rm int}$, the intervention indicator is set to zero. As such, and in accordance with Assumption \ref{assump1}, we do not use ${\bf Z}_{\rm int}$ within the model used to impute ${\bf Y}_{\rm int}$. Therefore, only the intermediate variables are used to construct the imputed values of the long-term outcome.

Using the imputed version of ${\bm \chi}$, the treatment effect $\alpha$ can be estimated from
\begin{equation} \label{outcomeeq}
E[Y_i|Z_i] = \mu + \alpha Z_i, 
\end{equation}
where $Y_i$ is the outcome and $Z_i$ is an indicator of treatment status for individual $i$. 
The imputation approach is valid for estimating the treatment effect in (\ref{potential}) since multiply imputing $Y_i$ will give a reasonable approximation to $E[{Y}_i|{X}_i={x}]$, and then averaging over cases in the respective treatment conditions will encapsulate the contributions of $f^{(1)}(x)$ and $f^{(0)}(x)$.

Similar to how ${\bf Z}_{\rm int}$ is not used in the imputation model for ${\bf Y}_{\rm int}$, we only use cases in the intervention dataset to estimate the parameters of the above model.  One could use all available cases in both datasets (while setting ${\bf Z}_{\rm int}=\bm{0}$), but this would require a stronger degree of congeniality between the two datasets: the marginal distribution of the intermediate and outcomes variables would need to be equivalent across both datasets.

\subsection{Variance Estimation}

Estimates of parameters like $\alpha$ in (\ref{outcomeeq}) are subject to uncertainty due not only to sampling error but also due to imputation error (as the imputed values cannot be assumed to be equal to the true value of the long-term outcome that would have been observed if possible). To account for imputation error when approximating the variance inherent in relevant estimators, we borrow ideas from the multiple imputation framework.  Therein, $m$ (for $m>1$) imputed versions of $\bm{\chi}$ are created independently of one another. In theory, the multiple imputed datasets allow the analyst to gauge the magnitude of the imputation error by quantifying the variance in the imputed values across datasets.

\subsubsection{Combining rules for multiple imputation}

Inference for multiply imputed datasets is typically performed using algebraic rules which combine the datasets and adjust variances for imputation error. For a variety of reasons, these rules may not be applicable in our setting; however, we review them below.

Let $\widehat{\alpha}_k$ denote the estimated value of $\alpha$ from (\ref{outcomeeq}) that is estimated using the $k^{\rm th}$ dataset, and let $\hat\nu_k$ denote an estimator of the variance of $\widehat{\alpha}_k$.  The point estimator of $\alpha$ is
$
\bar{\alpha} = m^{-1}\sum^{m}_{k=1}\widehat{\alpha}_k,
$ 
and the within imputation and between imputation variances are
$ 
W_m = m^{-1}\sum^{m}_{k=1}\hat\nu_k
$
 and 
$
B_m = (m-1)^{-1}\sum^{m}_{k=1}(\widehat{\alpha}_k-\bar{\alpha} )^2,
$ 
respectively.

In the classic multiple imputation setting, the total variance of $\bar{\alpha}$ is calculated using
\begin{equation} \label{mi}
T_{\rm mi}=W_m+(1+1/m)B_m.
\end{equation}  
However, we do not expect this formula to be applicable in our setting because only a portion of the dataset that used for imputation is also used for analysis (i.e., most of the outcomes dataset is used to generate the imputations but is disregarded when estimating the treatment effect). This invalidates the classic multiple imputation estimate of the total variance.

For alternative combining rules, we draw on the domain of data confidentiality, wherein a fully imputed dataset is created by sampling from the posterior distribution of the observed data. The analogue in our setting is that the outcomes dataset conditional on the intermediate variables represents the observed data, and the imputed outcomes conditional on the intermediate variables represent the synthetic data. \cite{raghunathan03} first broached this problem and suggest a variance estimator of
\begin{equation} \label{rag}
T_{\rm syn}=(1+1/m)B_m-W_m.
\end{equation}
This estimator has been shown to perform poorly in small samples \citep{little15}, and can sometimes lead to a negative estimate of the variance.

In light of issues surrounding $T_{\rm mi}$, \cite{raab16} outline alternatives, including
\begin{equation} \label{raab1}
T_{\rm PPD}=\left(\rho + \frac{1+\rho}{m}\right)W_m,
\end{equation}
when posterior predictive sampling is used, and
\begin{equation} \label{raab2}
T_{\rm s}=\left(\rho + \frac{1}{m}\right)W_m.
\end{equation}
when it is not, where $\rho$ is the ratio of the sample sizes of the synthetic and observed datasets. \cite{raab16} also suggest an estimator for situations involving partially synthetic data:
\begin{equation} \label{raab3}
T_{\rm p}=W_m + \frac{B_m}{m}.
\end{equation}

Like (\ref{mi}), Equations (\ref{rag})-(\ref{raab3}) were not designed for our specific setting. Primarily, our analysis of interest involves modeling the conditional relationship between long-term outcomes and a variable that is not used for imputation, the intervention indicator (${\bf Z}_{\rm int}$).  Simulations (see Section \ref{sims}) show that these formulas do not perform well in our setting.  

\subsubsection{Jackknife for multiple imputation} \label{jack}

As an alternative to the combining rules that are typically seen in multiple imputation and synthetic data settings, we consider replication methods for variance estimation.  Specifically, we examine a delete-$d$ (or delete-a-group) jackknife \citep{shao89, kott01}.
To apply this procedure, both the intervention and outcomes datasets ($\bm{\chi}_{\rm int}$ and $\bm{\chi}_{\rm out}$ from (\ref{chidat})) are segmented into $G$ mutually exclusive and exhaustive groups denoted $\bm{\chi}^{(g)}_{\rm int}$ and $\bm{\chi}^{(g)}_{\rm out}$ for $g = 1,\ldots,G$, so that $\bm{\chi}_{\rm int} = ((\bm{\chi}^{(1)}_{\rm int})';\ldots;(\bm{\chi}^{(G)}_{\rm int})')'$ and $\bm{\chi}_{\rm out} = ((\bm{\chi}^{(1)}_{\rm out})';\ldots;(\bm{\chi}^{(G)}_{\rm out})')'$. 
Furthermore, let the concatenated dataset $\bm{\chi}^{(g)} = ((\bm{\chi}^{(g)}_{\rm out})'; (\bm{\chi}^{(g)}_{\rm int})')'$ denote a version of (\ref{chidat}) containing only cases in the $g^{\rm th}$ group. The replicate datasets are then notated $\bm{\chi}^{(-g)}$, which is equivalent to $\bm{\chi}$ with cases in $\bm{\chi}^{(g)}$ removed.

We let $\theta$ denote a parameter of interest (e.g., $\alpha$ from (\ref{outcomeeq})) that may be a function of the unobserved values in ${\bf Y}_{\rm int}$,
and as such can be estimated using $\bar\theta = E[\theta|\bm{\chi}]$. It follows that
\[
\bar\theta = \frac{1}{G}\sum^{G}_{g=1}E[\theta|\bm{\chi}^{(g)}] = \frac{1}{G}\sum^{G}_{g=1}\theta^{(g)}
\]
where $\theta^{(g)}=E[\theta|\bm{\chi}^{(g)}]$ denote the so-called pseudovalues.  Consequentially, $\mbox{Var}(\bar\theta)$ can be approximated with a sample variance or, perhaps more efficiently, a jackknife wherein
\[
\widetilde{\mbox{Var}}(\bar\theta) = \frac{G-1}{G}\sum^{G}_{g=1}(\theta^{(-g)}-\bar\theta)^2,
\]
for $\theta^{(-g)}=E[\theta|\bm{\chi}^{(-g)}]$ so that $\bar\theta = (1/G)\sum^{G}_{g=1}\theta^{(-g)}$. Note that the pseudovalues can be expressed $\theta^{(g)} = G\bar\theta - (G-1)\theta^{(-g)}$.

The applicability of a jackknife with imputed data has been examined previously \citep{rao92, righi14}; these authors recommend single imputation within each replicate group. However, they address the limited circumstance of hot deck imputation across a single imputed variable. To extend the applicability of the jackknife to more general settings, we consider multiply imputing within each replicate group; as such, we propose using multiple imputation to determine the value of $\theta^{(-g)}$.

To elaborate, for each $g = 1,\ldots, G$, the dataset $\bm{\chi}^{(-g)}$ is independently imputed $m$ times.  We let $\hat\theta^{(-g)}_j$ indicate the value of $\theta$ estimated from the $j^{\rm th}$ imputed version of $\bm{\chi}^{(-g)}$.  As such, our estimate of $\theta^{(-g)}$ is
$ 
\hat\theta^{(-g)} = m^{-1} \sum^{m}_{j=1} \hat\theta^{(-g)}_j,
$ 
and the jackknife estimate of $\theta$ is
\begin{equation} \label{meanG}
\bar\theta_{\rm jack} = \frac{1}{G} \sum^{G}_{g=1} \hat\theta^{(-g)},
\end{equation}
the variance of which is approximated using
\begin{equation} \label{varG}
\widehat{\mbox{Var}}(\bar\theta_{\rm jack}) = \frac{G-1}{G}\sum^{G}_{g=1}(\hat\theta^{(-g)}-\bar\theta_{\rm jack})^2.
\end{equation}

Note that the use of finite $m$ within each replication group will lead to some degree of upward bias in the variance estimates. Specifically, the validity of the jackknife is based upon the fact that the sample variance of the pseudovalues ($\theta^{(g)}$) then divided by $G$ gives a reasonable estimate of the variance of the desired quantity ($\bar\theta$). However, since $\theta^{(-g)}$ must be estimated, mean zero noise is added to the pseudovalues. That is, $\hat\theta^{(-g)} = \theta^{(-g)} + \epsilon^{(-g)}$, where $\mbox{Var}(\epsilon^{(-g)}) = Gc^2/[(G-1)m]$ for some constant term $c^2$ that is on the order of $\mbox{Var}(\bar\theta)$. This, in turn, implies $\hat\theta^{(g)} = \theta^{(g)} + \epsilon^{(g)}$ with $\mbox{Var}(\epsilon^{(g)})= G(G^2-3G+3)c^2/[m(G-1)]$.  Hence,
\[
\widehat{\mbox{Var}}(\bar\theta_{\rm jack}) = G^{-1}S^2_{\hat\theta \hat\theta} = \widetilde{\mbox{Var}}(\bar\theta) + 2G^{-1}S_{\theta \epsilon} + G^{-1}S^2_{\epsilon \epsilon}
\]
where $S^2_{\hat\theta \hat\theta}$ is the sample variance of the $\hat\theta^{(g)}$ for $g = 1,\ldots, G$, $S_{\theta \epsilon}$ is the sample covariance of the $\theta^{(g)}$ and the $\epsilon^{(g)}$, and $S^2_{\epsilon \epsilon}$ is the sample variance of the $\epsilon^{(g)}$. Note that $S_{\theta \epsilon}$ is negligible in comparison to $S^2_{\epsilon \epsilon}$ due to independence of $\theta^{(g)}$ and $\epsilon^{(g)}$, and it holds that
\[
G^{-1}S^2_{\epsilon \epsilon} \approx Gc^2/m.
\]
As such, $m$ needs to be large in comparison to $G$ in order to render $G^{-1}S^2_{\epsilon \epsilon}$ negligible. That is, $\widehat{\mbox{Var}}(\bar\theta_{\rm jack})$ diverges from $\widetilde{\mbox{Var}}(\bar\theta)$ with increasing $G$ for fixed $m$. 
In addition, $\bar\theta_{\rm jack} = \bar\theta + \bar\epsilon$, where $\mbox{Var}(\bar\epsilon) = c^2/(mG)$, which implies $\bar\theta_{\rm jack}$ converges to $\bar\theta$ with increasing $G$.
Consequentially, the jackknife procedure will fail if single imputation \citep[$m=1$, as recommended by][p.~83]{little02} is used.

\subsubsection{Bootstrap for multiple imputation} \label{boot}

An alternative to the jackknife is the bootstrap \citep{efron94}. Let $S^{(b)}$ denote a random sample, taken with replacement, of rows from $\bm{\chi}$. The cardinality of $S^{(b)}$ equals the total number of rows of $\bm{\chi}$ from (\ref{chidat}).  Let $\bm{\chi}^{(b)}$ denote a version of $\bm{\chi}$ containing only cases in $S^{(b)}$ (with repeated rows as needed), and let $\theta^{(b)}=E[\theta|\bm{\chi}^{(g)}]$ (treating repeated rows as unique observations). Assume that $B$ bootstrap samples are independently created, and let
\begin{equation} \label{meanB}
\bar\theta_{\rm boot} = B^{-1}\sum^{B}_{b=1}\theta^{(b)}.
\end{equation}
Ideally, the variance of $\bar\theta_{\rm boot}$ would be approximated with
\[
\widetilde{\mbox{Var}}(\bar\theta_{\rm boot}) = \frac{1}{B-1} \sum^{B}_{b=1} (\theta^{(b)} - \bar\theta_{\rm boot})^2;
\]
however, the values of $\theta^{(b)}$ are unknown.

Akin to our approach with the jackknife, and in accordance with the observations of \cite{schomaker18}, we recommend multiply imputing each of the bootstrap samples $m$ times in order to adequately approximate each $\theta^{(b)}$.  Let $\hat\theta^{(b)}_j$ denote the estimate of $\theta$ collected from the $j^{\rm th}$ imputed dataset for the $b^{\rm th}$ boostrapped sample, and therefore $\hat\theta^{(b)}=m^{-1}\sum^{m}_{j=1}\theta^{(b)}_j$ indicates the best estimator of $\theta^{(b)}$.  The bootsrapped estimate of $\theta$ is then $\hat\theta_{\rm boot}=B^{-1}\sum^{B}_{b=1}\hat\theta^{(b)}$, and the variance of $\hat\theta_{\rm boot}$ is approximated with
\begin{equation} \label{varB}
\widehat{\mbox{Var}}(\hat\theta_{\rm boot}) = \frac{1}{B-1} \sum^{B}_{b=1} (\hat\theta^{(b)} - \hat\theta_{\rm boot})^2.
\end{equation}

Note that $\hat\theta^{(b)} = \bar\theta^{(b)} + \epsilon^{(b)}$ where $E[\epsilon^{(b)}]=c^2/m$ and $\mbox{Var}(\epsilon^{(b)})=c^2/m$.  It follows that
\[
\widehat{\mbox{Var}}(\hat\theta_{\rm boot}) = \widetilde{\mbox{Var}}(\bar\theta_{\rm boot}) + 2V_{\theta \epsilon} + V^2_{\epsilon \epsilon},
\]
where  $V_{\theta \epsilon}$ is the sample covariance of the $\theta^{(b)}$ and the $\epsilon^{(b)}$, and $V^2_{\epsilon \epsilon}$ is the sample variance of the $\epsilon^{(g)}$. As before, $V_{\theta \epsilon}$ is negligible in comparison to $V^2_{\epsilon \epsilon}$. Furthermore, $V^2_{\epsilon \epsilon}\approx c^2/m$, and as such the discrepancy between $\widehat{\mbox{Var}}(\hat\theta_{\rm boot})$ and $\widetilde{\mbox{Var}}(\bar\theta_{\rm boot})$ decreases with increasing $m$ and in effect does not depend upon $B$.

The bootstrap will require more bootstrapped samples ($B$) than the jackknife requires replication groups ($G$); however, since (unlike the jackknife) $\hat\theta^{(b)}$ is not divergent from $\theta^{(b)}$ with increasing $B$ for fixed $m$, the bootstrap may not require $m$ to be as large as is needed with the jackknife.

One could conceivably estimate $E[\theta|\bm{\chi}]$ (and $E[\theta|\bm{\chi}^{(g)}]$ or $E[\theta|\bm{\chi}^{(b)}]$) through means other than imputation. 
However, we prefer imputation since algorithms such as \texttt{gerbil} provide a straightforward manner to handle data of general structures (e.g., continuous, binary, categorical, etc.). Furthermore, imputation provides a natural manner through which missing values in the intermediate variables can be handled (i.e., those will be imputed in tandem with each imputed version of ${\bf Y}_{\rm int}$), and through which multiple intermediate variables (and multiple outcome variables) can be simultaneously handled.

\section{Simulations} \label{sims}

Here, we present a simulation study to evaluate the efficacy of the methods outlined in Section \ref{sec2}. We generate synthetic versions of intervention and outcomes datasets.  We create 4 intermediate variables (${X}_1,\ldots,{X}_4$) as follows.  We first sample a 4-dimensional vector, $\bm{W} = (W_1, \ldots, W_4)'$, from a multivariate normal distribution with a mean vector of 0 and a covariance matrix with 1 on the diagonals and 0.5 on each off-diagonal. The first $n_{\rm int}$ observations are considered part of the intervention dataset. Each observation in the intervention dataset is then assigned a treatment indicator $Z$ which takes on a value of one with probability $p = 0.5$ and a value of zero otherwise.  For cases in the interventions dataset, we generate the observed intermediate variables via $X_1 = W_1$, $X_2 = W_2$, $X_3 = W_3 + 0.5Z$, and $X_4 = W_4 + 0.5Z$. That is, only the third and fourth intermediate variable are affected by the treatment. However, for cases in the outcomes dataset, we set $X_j = W_j$ for $j = 1,\ldots,4$.

For all cases, we generate the outcome is generated as follows:
\[
Y = 0.5 X_1 + 0.5 X_2 + 0.5 X_3 + 0.5 X_4 + \epsilon,
\]
where $\epsilon \sim N(0,1)$. To account for the possibility that only a segment of a population will be eligible for the intervention (as is the case in our applications), we drop cases with $X_1 < -0.5$ from the simulated interventions dataset.  Following these exclusions, it holds that $n_{\rm int} = 500$ on average with $n_{\rm out} = 500$ in this setup.

Next, we consider the values of $Y$ to be missing for the outcomes dataset and impute them with an imputation model that uses the concatenated dataset with the outcome modeled on $X_1,\ldots,X_4$ but not $Z$.  Estimation of the treatment effect is performed by fitting the model in (\ref{outcomeeq}) 
for only cases in the interventions dataset while using linear regression. First, we use multiple imputation with $m=200$ and with combining rules (\ref{mi})-(\ref{raab3}) to estimate variance. In addition, we implement a jackknife with $G=25$ and various values of $m$ and use (\ref{meanG}) and (\ref{varG}) to estimate mean and variance of the treatment effect, respectively, and we consider and a bootstrap with $B=250$ and various values of $m$ while using (\ref{meanB}) and (\ref{varB}) to estimate the mean and variance, respectively.

The above data generation process is repeated $R = 2000$ times. For a given method, we let $\hat\theta^{[r]}$ denote the point estimate of a parameter $\theta$ estimated at the $r^{\rm th}$ replication.
For each method, we calculate the bias and root mean-squared error (rMSE) in the estimate of $\theta$ as follows:
\[
\mbox{bias}_\theta = \frac{1}{R}\sum^{R}_{r=1}(\hat\theta^{[r]}-\theta),
~~~~~~\mbox{and}~~~~
\mbox{rMSE}_\theta = \sqrt{\frac{1}{R}\sum^{R}_{r=1}(\hat\theta^{[r]}-\theta)^2}.
\]
For each method, we denote the corresponding estimate for $\mbox{Var}(\hat\theta^{[r]})$ with $\hat{V}^{[r]}$. The $100(1-\gamma)^{\rm th}$ confidence interval for $\theta$ is then calculated via
\[
C_{1-\gamma} = \{
x: x \geq \hat\theta^{[r]} - z_{1-\gamma/2} \sqrt{\hat{V}^{[r]}} ~~ \& ~~ x \leq \hat\theta^{[r]} + z_{1-\gamma/2}\sqrt{\hat{V}^{[r]}}
\},
\]
where $z_{1-\gamma/2}$ is the $100(1-\gamma)^{\rm th}$ percentile of the standard normal distribution. (Since $m$ is large, we do not worry about a correction for finite $m$ that would mandate a $t$-distribution for the multiple imputation confidence intervals.) The coverage for each method is approximated using
$
\mbox{cover}_{\theta}=R^{-1}\sum^{R}_{r=1}\bm{1}_{\theta \in C_{1-\gamma}},
$ 
which indicates the portion of replications in which the confidence interval contains the true value of $\theta$.

Results for the for the simulation process outlined above (referred to as the primary data fusion model) are shown in Table \ref{tab1}.  Therein, we first of observe that none of the methods based on multiple imputation combining rules perform well in terms of estimated coverage (although, as expected, multiple imputation does yield an unbiased point estimator).  Furthermore, we see that, for sufficiently large $m$, both the jackknife and bootstrap methods yield unbiased estimates and appropriate coverage levels for both the intercept and treatment effect parameters. Furthermore, we see that for smaller $m$, these two methods give conservative intervals, implying, as theorized in Section \ref{sec2}, that variance is overstated in those cases.  As also theorized in Section \ref{sec2}, it appears the jackknife requires a substantially larger value of $m$ to achieve sufficient coverage than is required by the bootstrap.

\begin{table}[!ht]
\caption{\label{tab1}Simulated bias, rMSE, and coverage for the intercept and treatment effect parameters from (\ref{outcomeeq}) when using different methods (i.e., multiple imputation combining rules from (\ref{mi})-(\ref{raab3}), jackknife with various $m$, and bootstrap with various $m$).
\vspace{.08in}
}
\centering
\setlength{\tabcolsep}{.4em}
\begin{tabular}{clccccccc}
\hline \hline
& & \multicolumn{3}{c}{Intercept ($\mu$)} & & \multicolumn{3}{c}{Treatment effect ($\alpha$)} \\ \cline{3-5} \cline{7-9}
& & Bias & rMSE & Cover. & & Bias & rMSE & Cover. \\  \hline
\multirow{5}{*}{\specialcell{Combining \\ Rules}} & $T_{\rm mi}$ from (\ref{mi}) & -0.0004 & 0.0957 & 0.9920 &  & 0.0017 & 0.1240 & 0.9930 \\
& $T_{\rm syn}$ from (\ref{rag}) & -0.0004 & 0.0957 & --- &  & 0.0017 & 0.1240 & --- \\
& $T_{\rm PPD}$ from (\ref{raab1}) & -0.0004 & 0.0957 & 0.9710 &  & 0.0017 & 0.1240 & 0.9835 \\
& $T_{\rm s}$ from (\ref{raab2}) & -0.0004 & 0.0957 & 0.9720 &  & 0.0017 & 0.1240 & 0.9835 \\
& $T_{\rm p}$ from (\ref{raab3}) & -0.0004 & 0.0957 & 0.9710 &  & 0.0017 & 0.1240 & 0.9835 \\ \hline
\multirow{7}{*}{\specialcell{Jackknife \\ ($G=25$)}} & $m = 1$ & -0.0000 & 0.0968 & 1.0000 &  & 0.0015 & 0.1254 & 1.0000 \\
& $m = 5$ & -0.0004 & 0.0958 & 0.9990 &  & 0.0019 & 0.1244 & 0.9985 \\
& $m = 10$ & -0.0005 & 0.0957 & 0.9955 &  & 0.0018 & 0.1243 & 0.9945 \\
& $m = 25$ & -0.0004 & 0.0954 & 0.9830 &  & 0.0016 & 0.1242 & 0.9825 \\
& $m = 50$ & -0.0005 & 0.0953 & 0.9700 &  & 0.0017 & 0.1241 & 0.9670 \\
& $m = 100$ & -0.0005 & 0.0953 & 0.9565 &  & 0.0017 & 0.1241 & 0.9575 \\
& $m = 200$ & -0.0005 & 0.0953 & 0.9495 &  & 0.0016 & 0.1241 & 0.9515 \\ \hline
\multirow{4}{*}{\specialcell{Bootstrap \\ ($B=250$)}} & $m = 1$ & -0.0005 & 0.0955 & 0.9895 &  & 0.0016 & 0.1244 & 0.9870 \\
& $m = 5$ & -0.0003 & 0.0955 & 0.9690 &  & 0.0014 & 0.1242 & 0.9610 \\
& $m = 10$ & -0.0003 & 0.0954 & 0.9620 &  & 0.0014 & 0.1242 & 0.9545 \\
& $m = 25$ & -0.0003 & 0.0954 & 0.9580 &  & 0.0014 & 0.1242 & 0.9525 \\  \hline
\end{tabular} 
\end{table}

To assess the performance of data fusion when the ideal setup (in terms of intermediate variables and data units) is not used, we consider three separate data fusion models as sensitivity analyses.  These include:
\begin{enumerate}
\item \underline{Reduced Outcomes Dataset}: Same as the primary model analyzed in Table 1 with the exception that the cases used from the outcomes model are restricted so that the outcomes dataset best resembles the intervention dataset.  That is, the imputation process only includes cases from the outcomes dataset which observed $X_1 < -0.5$.  This setting is designed to evaluate the benefits of aligning the two datasets.
\item \underline{Drop $X_3$}: Same as the primary model analyzed in Table 1 with the exception that one of the two variables affected by the intervention ($X_3$) is dropped from the imputation model. This is designed to introduce a departure from Assumption \ref{assump1} via a situation in which an unobserved confounder that is affected by the intervention exists (akin to Example 1 of Section \ref{examples}).
\item \underline{No Covariates}: Same as the primary model analyzed in Table 1 with the exception that only the two variables which are affected by the treatment ($X_3$ and $X_4$) are used as intermediate variables (that is, $X_1$ and $X_2$ are dropped from the imputation model). This is designed to introduce a departure from Assumption \ref{assump1} via a situation in which an unobserved confounder that is {\em not} affected by the intervention (akin to Example 2 of Section \ref{examples}).
\item \underline{Different Conditionals}: Same as the primary model analyzed in Table 1 with the exception that the outcome is generated via the equation $Y = 0.6 X_1 + 0.6 X_2 + 0.6 X_3 + 0.6 X_4 + \epsilon$ for cases in the outcomes dataset. This is designed to test the sensitivity to deviations from Assumption \ref{assump2}.
\end{enumerate}
Results for these sensitivity analyses are shown in Table \ref{tab2}.

\begin{table}[!ht]
\caption{\label{tab2}Simulated bias, rMSE, and coverage for the intercept and treatment effect parameters from (\ref{outcomeeq}) for three models used for sensitivity analysis. Only results for the jackknife (with $m=200$) and bootstrap (with $m=25$) are shown.
\vspace{.08in}
}
\centering
\setlength{\tabcolsep}{.4em}
\begin{tabular}{llccccccc}
\hline \hline
& & \multicolumn{3}{c}{Intercept ($\mu$)} & & \multicolumn{3}{c}{Treatment effect ($\alpha$)} \\ \cline{3-5} \cline{7-9}
& & Bias & rMSE & Cover. & & Bias & rMSE & Cover. \\  \hline
\multirow{2}{*}{{Reduced Outcome Dataset}} & Jackknife & 0.0031 & 0.0961 & 0.9570 &  & -0.0006 & 0.1259 & 0.9535 \\
& Bootstrap & 0.0030 & 0.0962 & 0.9600 &  & -0.0003 & 0.1262 & 0.9565 \\ \hline
\multirow{2}{*}{{Drop $X_3$}} & Jackknife & -0.0014 & 0.0978 & 0.9430 &  & -0.1873 & 0.2210 & 0.6530 \\
& Bootstrap & -0.0013 & 0.0979 & 0.9485 &  & -0.1875 & 0.2211 & 0.6440 \\ \hline
\multirow{2}{*}{{No Covariates}} & Jackknife & -0.2110 & 0.2327 & 0.4735 &  & 0.3332 & 0.3548 & 0.2535 \\
& Bootstrap & -0.2109 & 0.2329 & 0.4540 &  & 0.3331 & 0.3549 & 0.2330 \\ \hline
\multirow{2}{*}{{Different Conditionals}} & Jackknife & 0.1287 & 0.1693 & 0.8020 &  & 0.1010 & 0.1754 & 0.8985 \\
& Bootstrap & 0.1289 & 0.1696 & 0.8095 &  & 0.1004 & 0.1752 & 0.9040 \\  \hline
\end{tabular} 
\end{table}

From results for the ``Reduced Outcomes Dataset'' model in Table \ref{tab2}, we see that restricting the outcomes dataset to align with the interventions dataset also produces unbiased point estimates with interval estimates that have appropriate coverage.  However, the rMSE for the primary model (when the outcomes dataset is not restricted) is smaller than what is obtained in the ``Reduced Outcomes Dataset'' model with the restrictions (this implies that cases in the outcomes dataset that do not necessarily resemble cases in the interventions dataset may still be informative for producing reasonable imputations for the final outcomes in the interventions dataset.  Furthermore, the results for the ``Drop $X_3$'' model indicate a biased estimate of the treatment effect. This is expected as suggested by Example 1 in Section \ref{examples} since Assumption \ref{assump1} is violated in this case.  However, the ``No Covariates'' model shows that when an unobserved confounder exist that is not affected by the treatment, we still have the potential for a biased estimate of the treatment effect as is illustrated by Example 2 of Section \ref{examples}. Furthermore, we see that departures from Assumption \ref{assump2} may introduce bias as illustrated by the results for the ``Different Conditionals''  model in Table \ref{tab2}.



\section{Data for a Health Insurance Case Study}


In the following sections, we use data fusion to estimate the effect of health insurance on the long-term outcome of mortality.
%
Health insurance can impact morbidity and mortality in several ways.  Findings from the OHIE suggest that among lower-income populations, health insurance may improve the use of primary and preventive health care services \citep{baicker2013oregon} and increase the probability of emergency department visits \citep{finkelstein2012oregon, taubman2014oregon}.  It may also increase the likelihood of diagnoses for depression and chronic conditions such as diabetes,  and increase use of prescription drugs \citep{baicker2013oregon}.  Health insurance can improve mental and physical health outcomes by reducing stress and financial hardship due to high and uncertain medical expenses \citep{finkelstein2012oregon, baicker2013oregon}.  Overall, lottery winners in the OHIE reported better self-reported physical and mental health within the first two years of the lottery, though there were no changes in clinical measures of physical health, such as blood pressure and cardiovascular risk factors \citep{finkelstein2012oregon}.

One challenge in evaluating the impact of health insurance on mortality is that health is a stock variable and responds slowly to improved access and care.  As a result, long-term data is usually required to yield evidence of effects on mortality.
This practical difficulty of estimating long-term impacts of insurance provides a primary rationale for using data fusion methods.

First, we describe the datasets used in this process: The quasi-experimental Oregon Health Insurance Experiment (OHIE) is used as the intervention data that and the National Longitudinal Mortality Study (NLMS) is used as the outcomes dataset.

\subsection{The Oregon Health Insurance Experiment}
\label{ohie}


In early 2008, Oregon opened a wait list for 10,000 slots in the Oregon Health Plan (OHP) Standard, the state Medicaid program for low-income adults who are not categorically eligible for a second Medicaid plan (OHP Plus).\footnote{ This section draws heavily on \cite{finkelstein2012oregon} and \cite{baicker2013oregon}.  These studies provide additional detail on the Oregon experiment and prior evaluations.}  After receiving close to 90,000 applicants, the state used a series of lottery drawings to determine who would have the opportunity to apply for OHP Standard, subject to submitting an application and meeting the eligibility criteria.  Ultimately 30\% of the selected individuals successfully enrolled in the health insurance program.  The eligible population for OHP Standard includes adults 19-64 who are not eligible for other public insurance programs, and have been without health insurance in the past six months, as well as income below the Federal Poverty Level (FPL; equivalent to annual income of \$30,657 for a family of four) and assets below \$2,000.  The benefits are comprehensive and provided without cost-sharing.  Coverage is at the household level.  Enrollees are responsible for premiums between \$0-20 per month, depending on their income.

As noted above, the OHIE has made several important contributions to understanding the impact of health insurance within one to two years after the lottery.  However, the effect of the lottery on the long-term outcome considered here has not been examined in existing OHIE studies.

Of course, our findings from data fusion involving the OHIE are contingent upon Assumptions \ref{assump1} and \ref{assump2} holding. Although these assumptions cannot be empirically verified, we argue that these analyses are very well-suited for them to hold. That is, it is hard to conceive that the OHIE lottery would have a direct effect on the long-term outcomes that could not be quantified through intermediate variables that are measured. For example, winning the lottery should only directly affect whether or not one has health insurance, and possession of health insurance is an intermediate variable that is captured in all datasets considered here (in addition, hopefully, to all confounders that may correlate with both this variable and with the long-term outcomes have been captured as covariates here). In addition, we consider the robustness of our findings to the inclusion of this intermediate variable.

\subsection{The National Longitudinal Mortality Study}
\label{NLMS}

We combine the OHIE data on participants approximately one year after the lottery (2010) with data from the National Longitudinal Mortality Study \citep[NLMS;][]{sorlie95, NLMS14} to estimate the effect of the health insurance lottery on mortality.


Our analyses involve the NLMS 11-year follow up file, which consists of 1,835,072 individuals that were followed prospectively for 11 years after an initial interview. Each record in the file includes demographic and socioeconomic variables collected at the time of the interview combined with a mortality outcome, if there is one. The 11-year file is weighted to match the US population in 1990, although it is not the case that all individuals in the file were interviewed at that time.

The covariates (which are not affected by the intervention) and the intermediate variables (which are affected by the intervention) used in the fusion of the OHIE and NLMS are seen in Table \ref{vardescr9}---these variables are available in both the OHIE and NLMS data. Note that the intermediate variables include \texttt{insur\_any} (the respondent has health insurance during the year prior to the interview) and \texttt{health} (self-reported overall health). The outcome variable is 11-year mortality and is only available in the NLMS.

The analytic dataset used for data fusion is produced by including cases that have a non-missing value for at least one of the two intermediate variables and a reported age between 18 and 65. After applying these criteria, the Oregon dataset has 31,377 cases (of which 95.2\% have a reported household income that is no more than twice the federal poverty line), and the NLMS has 763,495 cases (of which 27.9\% have a reported income of no more than twice the federal poverty line). 
Of the cases in our Oregon dataset, 15,594 are considered treated (i.e., were selected for the lottery) and 15,783 were not.

Some variables have missingness inherent within the analytic datasets.  The rates of missingness are given in Table \ref{bigtablea8} for subjects from the NLMS and the two cohorts (lottery winners and non-lottery winners) of the Oregon study.   Further, means and standard deviations of all variables across these groups are reported in Table \ref{bigtablea8}.  Only observed values are used in calculation of the means and standard deviations.  The table indicates the influence of the OHIE lottery on insurance status and self-reported health.  
The table also illustrates the similarities and disparities between the respondents from the NLMS and Oregon studies.

Also of interest are the relationships between variables.  Table \ref{bigtable1a8} shows pairwise correlations between the two intermediate variables and each variable considered in this study when calculated using respondents from the three cohorts (NLMS, Oregon lottery winners, and Oregon lottery non-winners).

\section{Fusion of OHIE and NLMS data}

In order to estimate the influence of the Oregon Medicaid lottery on longer-term mortality, we generate imputations of 11-year mortality events for individuals in the Oregon data.  We concatenate the Oregon and NLMS datasets, linking across variables that they have in common. We do not use a lottery status indicator in the concatenated dataset. The outcome variable (11-year mortality) is set to missing for cases coming from the Oregon data and is imputed using the R package \texttt{gerbil} \citep{robbins20, robbins21} while using the covariates and intermediate variables as predictors within the imputation model (although missingness in these variables is imputed in tandem with the outcomes).  Within the imputation process, an empirical distribution transformation \citep{robbins14} is applied to continuous variables. 


Following the simultaneous imputation of 11-year mortality for all cases in the Oregon dataset, we apply the treatment effect model of (\ref{outcomeeq}) to the imputed mortality events for the Oregon dataset. (Note that this model incorporates the treatment, i.e., lottery, status indicator which was not used within the imputation process.) This model is fit using linear regression and as such, the treatment effect is the difference between mean imputed mortality rate for the lottery and the corresponding rate for the non-lottery groups.

Note that estimation is performed using the jackknife procedure described in Section \ref{jack} with $G=25$ replication groups and $m=200$ imputed datasets per replication group. The jackknife is chosen in lieu of a bootstrap because the former was shown to yield slightly better rMSE in the simulations of Section \ref{sims}. For thoroughness, the data fusion process is repeated for four sensitivity analyses that involve the use of different inclusion criteria, intermediate variables, or covariates.
To elaborate, the primary model and sensitivity models are outlined as follows:

\begin{itemize}
\item \underline{Primary}:~All variables from Table \ref{vardescr9} are used in the data fusion process, and all cases that have an age between 18 and 65 years with an observed value of at least one of \texttt{insur\_any} and \texttt{health} are included.
\item \underline{Reduced NLMS}:~Same as the primary model, but efforts are taken to align the two datasets by applying data fusion to only cases that have a household income of no more than twice the federal poverty line (as such, 95\% of the Oregon dataset and 28\% of the NLMS are used).
\item \underline{Drop Insurance}:~Same as the primary model, but the intermediate variable that indicates whether the respondent has insurance at follow-up (\texttt{insur\_any}) is excluded from the data fusion process, and the restriction is imposed that a case must have an observed value of self reported health (\texttt{health}) in order to be included (as such, 99\% of the Oregon dataset and 28\% of the NLMS are used). 
\item \underline{Drop Health}:~Same as the primary model, but the intermediate variable that indicates self-reported health (\texttt{health}) is excluded from the data fusion process.
\item \underline{No Covariates}:~Same as the primary model, but the covariates (i.e., the variables listed in Table \ref{vardescr9} that labeled with type ``Covariate'' and thus assumed to not be affected by the intervention) are excluded from the data fusion process.
\end{itemize}

Aside from the restrictions on variables incorporated, the imputation model for each analysis enables all feasible dependencies.

\begin{figure}[ht!]
\centering
\begin{tabular}{cc}
\includegraphics[height=3.1in, bb = 8 8 462 426]{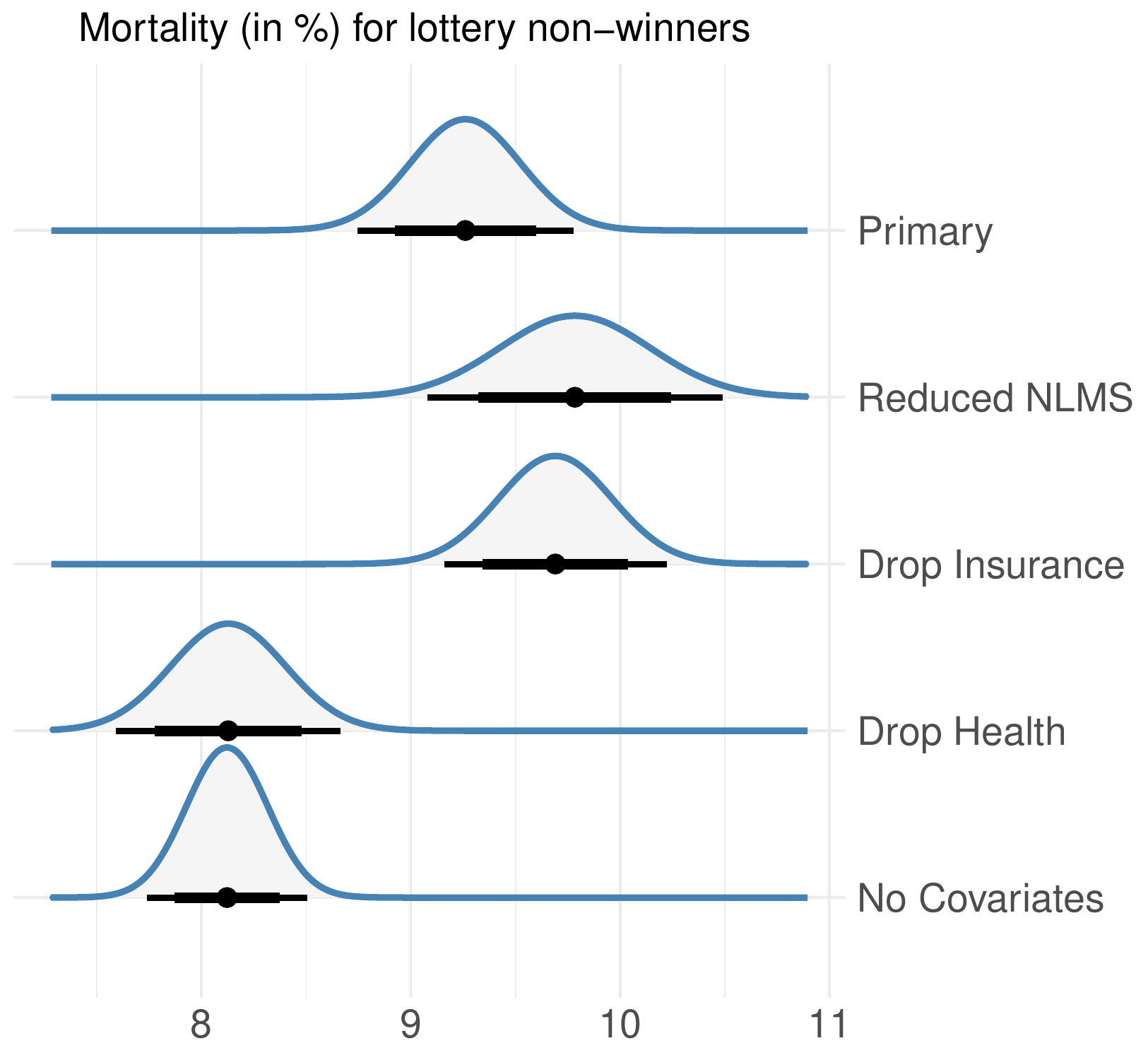}
&
\hspace{-.24in}
\includegraphics[height=3.1in, bb =  8 8 426 426]{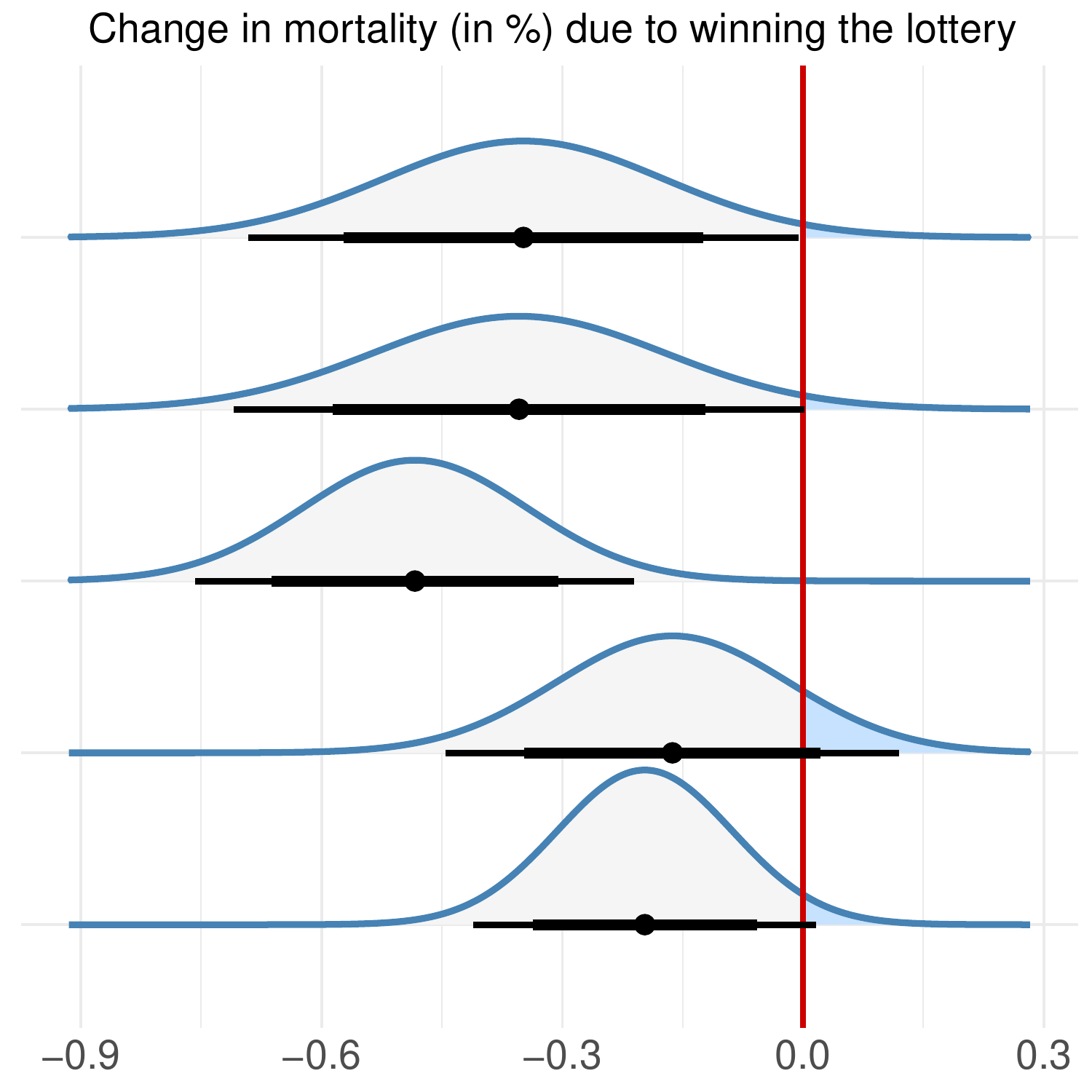} \\
\end{tabular}
\caption{Plots of the sampling distribution overlaid on respective 95\% confidence intervals for estimates of the intercept (left) and treatment (right) parameters from (\ref{outcomeeq}) for the 11-year mortality outcome as found using imputations generated via the various OHIE/NLMS data fusion models.}
\label{ohienlms}
\end{figure}

The results of the primary data fusion model are shown in Figure \ref{ohienlms}. The figure shows the estimated value of the intercept (the parameter $\mu$ from (\ref{outcomeeq})) and the treatment effect ($\alpha$), along with corresponding sampling distributions overlaid on respective 95\% confidence intervals, for the various models of the effect of the lottery on the imputed value of 11-year mortality. The results for the primary model indicate that winning the Oregon lottery was associated with a decrease in 11-year mortality. Specifically, those not winning the lottery had a 9.26\% chance of reporting a death event in the 11 years after the initial follow-up interview, whereas this probability is smaller by 0.35\% (95\% CI: 0.01\% to 0.69\%) for individuals that win the lottery.

The sensitivity analyses illustrate similar findings to the primary analysis.  First, the reduction of the datasets (mainly the NLMS) to include cases that have an income that is at most double the federal poverty line (the ``Reduce NLMS'' model) yields an estimated treatment effect of 0.35\% (95\% CI: 0.00\% to 0.71\%), which is nearly identical to the primary model.  (Note that the ``Reduced NLMS'' model yields a confidence interval that is slightly wider than the primary model, which is expected in accordance with Model 1 of Section \ref{sims}.)  However, if covariates are excluded from the imputation model (the ``No Covariates'' model), the treatment effect diminishes to a 0.20\% (95\% CI: -0.02\% to 0.41\%) decrease in mortality. This latter finding is unsurprising as, in accordance with Example 3 in Section \ref{examples}, the exclusion of covariates that relate to both the intermediate variables and the final outcome may lead to bias in the treatment effect.

Exclusion of one of the intermediate variables is shown to have some impact on the estimated treatment effect.  In particular, the treatment effect is enhanced to a 0.48\% (95\% CI: 0.21\% to 0.76\%) decrease in 11-year mortality when self-reported health is the only intermediate variable included in the imputation model (the ``Drop Insurance'' model) and drops to a 0.16\% (95\% CI: -0.12\% to 0.45\%) decrease in 11-year mortality when a indicator of health insurance is the only intermediate variable (the ``Drop Health'' model). These results are in line with the observation that although both intermediate variables are highly impacted by the lottery in the short-term, self-reported health is more correlated with 11-year mortality than is the indicator of health insurance (0.176 vs.~-0.004; see Table \ref{bigtable1a8}).

Note that although about 5\% of the NLMS cohort used in the primary model report a death event, our imputation model estimates that over 9\% of the OHIE control group will report a death event within 11 years of the initial OHIE follow-up. This observation illustrates the capability of the data fusion process to account for inherent discrepancies between the intervention and outcome datasets.

\section{Discussion}

The proposed data fusion approach addresses several issues of existing methods such as the product-of-estimates technique by handling for non-linear modeling with multiple intermediate variables, and we were able to establish a rigorous framework for making statistical inferences.
We have used this approach to estimate the long-term effects of health insurance on mortality, and we argued that the proposed methodology is well-suited for this application.  Prior to concluding, we offer several considerations for other researchers aiming to apply the techniques introduced here.

\subsection{Surrogate and Data Fusion Validity}

As noted throughout, the viability of data fusion for predicting long-term program impacts hinges upon a ``surrogate validity'' assumption which requires that the set of intermediate variables captures the effect of the treatment on the final outcome. Further, as shown in Example 2, confounding may lead to violation of this assumption even without the confounder being directly impacted by the intervention.  \cite{vanderweele2013surrogate} discusses departures from this assumption in the context of a ``surrogate paradox'' wherein it is argued that such issues may lead to a treatment negatively impacting a final outcome while positively impacting surrogate outcomes.

Our advice to the practitioner in regards to this assumption is as follows: The assumption is not more restrictive than ingorability assumptions required for routinely used propensity score methods, non-response weighting adjustments, and missing data analyses in general. As such, the literature on those procedures can offer guidance here.  In particular, one should collect a highly robust set of intermediate variables, including pre-treatment variables that may not behave as surrogates.
Further, substantive experts can give guidance on the selection of covariates to make this assumption plausible.
%

The proposed methodology also depends upon a ``data fusion validity'' assumption which stipulates that the outcomes dataset can be used to model the mechanism from which the final outcome in the interventions dataset would be generated conditional on the set of intermediate variables. This assumption has not received as much attention in the literature as the surrogate validity assumption, and it cannot be verified empirically.  However, from this assumption we ascertain that the two datasets need not have the same marginal characteristics so long as the appropriate conditional distributions are congenial.
For example, OHIE data considered here are a more restrictive sample than the NLMS (primarily, OHIE applicants were notably poorer than the overall population). However, this in itself does not invalidate data fusion as shown here with simulations and sensitivity analyses. Additionally, OHIE data were measured two decades following the NLMS data, and it is possible that nationwide mortality rates, and causes of mortality, may have changed in that time period.  Even so, the conditional relationship between mortality and the intermediate variables may remain unchanged.

If assumption valdity remains a concern, the potential for bias can be assessed through sensitivity analyses to estimate treatment effects under various assumptions about the relationships between variables that are never observed jointly \citep[see, e.g.,][]{reiter2012bayesian}.

\subsection{Intervention Study Design}

Confidence in the findings from our techniques can be enhanced through an improved study design in future applications to (among other things) help ensure that both assumptions are satisfied.
Since the potential outcomes data sources will typically be long-running studies whose design cannot be altered, it is therefore up to the intervention study team to plan a study that can best take advantage of available resources and to optimize the breadth of intermediate variables that can be used.  Further, since even small differences in data definitions or measurement methods between two data sources can result in substantial analytic hurdles \citep[e.g.,][]{burgette2012nonparametric, schifeling19}, all efforts should be made in the intervention study's design stage to maximize the compatibility of the intervention and outcomes studies and, indeed, to assess whether an appropriate outcomes data source can be found.

%

It may be possible to find suitable outcomes data after the intervention data have been gathered. This was the case for the OHIE/NLMS analysis presented above.  However, these analyses would have been strengthened by a richer set of shared variables that those that happened to be collected in both studies.

\subsection{Conclusion}


We argue that, in many cases, data fusion methods can leverage existing longitudinal data sources to enable researchers to produce estimates of policy impacts that come closer to the gold-standard of long-term, randomized controlled trials.  Done properly, the intervention study should be designed with extant longitudinal data sources in mind so as to produce common variable definitions and to maximize the credibility of the data fusion models, as envisioned by conceptual models of substantive experts.  These extra planning steps entail negligible costs compared to the expense of running the long-term study and offer potential for substantial improvements in predicted long-term impacts of the intervention being studied.
%
However, as noted throughout, the findings hinge upon the validity of underlying assumptions which may be difficult to verify in practice.
Importantly, the data fusion approach makes these assumptions explicit and gives policymakers a concrete foundation on which to make important decisions.
That said, in light of ambiguity regarding assumption validity, we recommend that data fusion be used to supplement policymakers' understanding of the potential impacts of an intervention as opposed to a replacement for long-term studies.

Lastly, in our motivating intervention, the OHIE, the treatment was applied randomly.  In contrast, many evaluations will be of treatments that cannot be randomized due to logistical or ethical constraints.  We leave the cases of nonrandomized interventions for future work.


\setlength{\bibsep}{0.0pt}
\renewcommand{\bibfont}{\footnotesize}

\bibliographystyle{Chicago}
\bibliography{datFuseRef}





\begin{landscape}
\begin{table}[htp]
\centering
\begin{threeparttable}
\begin{tabular}{lllll}
Variable name & Description & Class & Available in & Type \\
\hline
\texttt{int} & Indicator of winning lottery & Binary & OHIE & Treatment\\
\texttt{age} & Age & Continuous & OHIE \& NLMS & Covariate \\
\texttt{hispanic} & Race:~hispanic & Binary & OHIE \& NLMS & Covariate \\
\texttt{white} & Race:~white & Binary & OHIE \& NLMS & Covariate \\
\texttt{black} & Race:~black & Binary & OHIE \& NLMS & Covariate \\
\texttt{edu} & Education & Ordinal & OHIE \& NLMS & Covariate \\
\specialcella{\texttt{povpct} \\ ~ } & \specialcella{Household income as a percentage of the federal \\ poverty line} & \specialcella{Continuous \\ ~} & \specialcella{OHIE \& NLMS \\ ~ } & \specialcella{Covariate \\ ~} \\
\texttt{employ} & Employed & Binary & OHIE \& NLMS  & Covariate \\
\texttt{male} & Gender is male & Binary & OHIE \& NLMS & Covariate \\
\texttt{hhsize} & Household size & Discrete & OHIE \& NLMS & Covariate \\
\texttt{insur\_any} & Has health insurance at any point in the past year & Binary & OHIE \& NLMS & Intermediate \\
\specialcella{\texttt{health} \\ ~ } & \specialcella{Self-rating of overall health \\ (1 = Excellent; ...; 5 = Poor)} & \specialcella{Ordinal \\ ~ } & \specialcella{OHIE \& NLMS \\ ~ } & \specialcella{Intermediate \\ ~ }  \\
\texttt{inddea} & 11-year mortality & Binary & NLMS & Outcome \\
\hline
\end{tabular}
\caption{Descriptions, classes, and types for variables used in the fusion of the OHIE and NLMS. Types of variables measured in both datasets include ``Covariate'' (i.e., those unaffected by the intervention) and ``Intermediate'' (i.e., those affected by the intervention''). NLMS variables are scaled to indicate measurement in 1990, and variables in the OHIE were measured in 2010. 
}
\label{vardescr9}
\end{threeparttable}
\end{table}
\end{landscape}

\clearpage\newpage

%

\begin{landscape}
\renewcommand*\rot{\multicolumn{1}{R{25}{.75em}}}
\begin{table}[htp]
\centering
\footnotesize
\begin{threeparttable}
{\small
\vspace{-.4in}
\begin{tabular}{clcccccccccccc}
& &
\rot{\texttt{inddea}} &
\rot{\texttt{health}} &
\rot{\texttt{insur\_any}} &
\rot{\texttt{age}} &
\rot{\texttt{hispanic}} &
\rot{\texttt{white}} &
\rot{\texttt{black}} &
\rot{\texttt{edu}} &
\rot{\texttt{povpct}} &
\rot{\texttt{employ}} &
\rot{\texttt{male}} &
\rot{\texttt{hhsize}}
   \\ \hline
\multirow{9}{*}{\begin{turn}{90}Rate Missing\end{turn}} & NLMS$^*$ & 0.000 & 0.717 & 0.000 & 0.000 & 0.003 & 0.003 & 0.003 & 0.000 & 0.000 & 0.005 & 0.000 & 0.000 \\
 & Control$^*$ & 1.000 & 0.009 & 0.008 & 0.000 & 0.013 & 0.013 & 0.013 & 0.010 & 0.188 & 0.146 & 0.000 & 0.166 \\
 & Treatment$^*$ & 1.000 & 0.007 & 0.007 & 0.000 & 0.013 & 0.013 & 0.013 & 0.008 & 0.198 & 0.152 & 0.000 & 0.174 \\
 & NLMS$^\dag$ & 0.000 & 0.716 & 0.000 & 0.000 & 0.003 & 0.003 & 0.003 & 0.000 & 0.000 & 0.006 & 0.000 & 0.000 \\
 & Control$^\dag$ & 1.000 & 0.009 & 0.009 & 0.000 & 0.012 & 0.012 & 0.012 & 0.010 & 0.000 & 0.007 & 0.000 & 0.000 \\
 & Treatment$^\dag$ & 1.000 & 0.008 & 0.007 & 0.000 & 0.014 & 0.014 & 0.014 & 0.008 & 0.000 & 0.006 & 0.000 & 0.000 \\
 & NLMS$^\ddag$ & 0.000 & 0.000 & 0.000 & 0.000 & 0.000 & 0.000 & 0.000 & 0.000 & 0.000 & 0.006 & 0.000 & 0.000 \\
 & Control$^\ddag$ & 1.000 & 0.000 & 0.008 & 0.000 & 0.012 & 0.012 & 0.012 & 0.009 & 0.189 & 0.146 & 0.000 & 0.167 \\
 & Treatment$^\ddag$ & 1.000 & 0.000 & 0.007 & 0.000 & 0.013 & 0.013 & 0.013 & 0.008 & 0.198 & 0.153 & 0.000 & 0.175 \\
\hline
\multirow{9}{*}{\begin{turn}{90}Means\end{turn}} & NLMS$^*$ & 0.049 & 2.16 & 0.83 & 39.80 & 0.12 & 0.76 & 0.09 & 2.50 & 385.00 & 0.74 & 0.48 & 3.24 \\
 & Control$^*$ & --- & 3.33 & 0.33 & 42.40 & 0.13 & 0.71 & 0.05 & 2.23 & 79.50 & 0.46 & 0.42 & 2.97 \\
 & Treatment$^*$ & --- & 3.23 & 0.49 & 42.30 & 0.13 & 0.70 & 0.05 & 2.24 & 81.20 & 0.47 & 0.43 & 3.07 \\
 & NLMS$^\dag$ & 0.064 & 2.55 & 0.66 & 37.80 & 0.20 & 0.61 & 0.15 & 2.01 & 113.00 & 0.56 & 0.42 & 3.57 \\
 & Control$^\dag$ & --- & 3.37 & 0.32 & 43.00 & 0.11 & 0.74 & 0.04 & 2.24 & 70.30 & 0.45 & 0.40 & 3.01 \\
 & Treatment$^\dag$ & --- & 3.25 & 0.50 & 42.80 & 0.12 & 0.73 & 0.03 & 2.25 & 71.90 & 0.47 & 0.41 & 3.12 \\
 & NLMS$^\ddag$ & 0.042 & 2.16 & 0.82 & 40.30 & 0.17 & 0.71 & 0.09 & 2.61 & 402.00 & 0.77 & 0.48 & 3.21 \\
 & Control$^\ddag$ & --- & 3.33 & 0.33 & 42.30 & 0.13 & 0.71 & 0.05 & 2.23 & 79.60 & 0.46 & 0.42 & 2.97 \\
 & Treatment$^\ddag$ & --- & 3.23 & 0.49 & 42.30 & 0.14 & 0.70 & 0.05 & 2.24 & 81.20 & 0.47 & 0.43 & 3.07 \\
\hline
\multirow{9}{*}{\begin{turn}{90}St.~Error\end{turn}} & NLMS$^*$ & 0.217 & 1.06 & 0.37 & 12.00 & 0.32 & 0.42 & 0.28 & 1.02 & 275.00 & 0.44 & 0.50 & 1.57 \\
 & Control$^*$ & --- & 1.05 & 0.47 & 12.10 & 0.33 & 0.46 & 0.22 & 0.87 & 64.60 & 0.50 & 0.49 & 1.88 \\
 & Treatment$^*$ & --- & 1.05 & 0.50 & 12.10 & 0.34 & 0.46 & 0.21 & 0.86 & 64.80 & 0.50 & 0.49 & 1.91 \\
 & NLMS$^\dag$ & 0.245 & 1.18 & 0.47 & 12.30 & 0.40 & 0.49 & 0.35 & 0.93 & 54.70 & 0.50 & 0.49 & 1.90 \\
 & Control$^\dag$ & --- & 1.04 & 0.46 & 12.10 & 0.31 & 0.44 & 0.19 & 0.85 & 49.30 & 0.50 & 0.49 & 1.91 \\
 & Treatment$^\dag$ & --- & 1.03 & 0.50 & 12.10 & 0.32 & 0.44 & 0.18 & 0.85 & 49.40 & 0.50 & 0.49 & 1.93 \\
 & NLMS$^\ddag$ & 0.200 & 1.06 & 0.38 & 11.60 & 0.37 & 0.46 & 0.28 & 1.01 & 290.00 & 0.42 & 0.50 & 1.57 \\
 & Control$^\ddag$ & --- & 1.05 & 0.47 & 12.10 & 0.34 & 0.46 & 0.22 & 0.87 & 64.70 & 0.50 & 0.49 & 1.89 \\
 & Treatment$^\ddag$ & --- & 1.05 & 0.50 & 12.10 & 0.34 & 0.46 & 0.21 & 0.86 & 64.90 & 0.50 & 0.49 & 1.91 \\
\hline
\end{tabular}
}
\caption{Missingness rates, means, and standard errors for the covariate and intermediate variables while calculated across various sets of study subjects as used in the OHIE/NLMS data fusion. Unless otherwise, noted, respondents between the ages of 18 and 64 with at least one of \texttt{health} and \texttt{insur\_any} observed are included. ``Control'' represents Oregon respondents who were not selected in the lottery, whereas ``Treatment'' represents Oregon
  respondents who were selected in the lottery.} \label{bigtablea8}
\begin{tablenotes}
\item[*] Only respondents eligible for the ``Primary'' OHIE/NLMS data fusion model are used in calculation of these values.  The same set of respondents also defines those eligible for the ``Drop Health'' and ``No Covariates'' models. 
\item[\dag] Only respondents eligible for the ``Reduce NLMS'' OHIE/NLMS data fusion model (i.e., those at or below 200\% the poverty line) are used in calculation of these values.  
\item[\ddag] Only respondents eligible for the ``Drop Insurance'' OHIE/NLMS data fusion model (i.e., those without a missing value of \texttt{health}) are used in calculation of these values. 
\end{tablenotes}
\end{threeparttable}
\end{table}
\end{landscape}

\begin{landscape}
\renewcommand*\rot{\multicolumn{1}{R{25}{.75em}}}
\begin{table}[htp]
\centering
\begin{threeparttable}
{
\begin{tabular}{clccccccccccccc}
& &
\rot{\texttt{inddea}} &
\rot{\texttt{health}} &
\rot{\texttt{insur\_any}} &
\rot{\texttt{age}} &
\rot{\texttt{hispanic}} &
\rot{\texttt{white}} &
\rot{\texttt{black}} &
\rot{\texttt{edu}} &
\rot{\texttt{povpct}} &
\rot{\texttt{employ}} &
\rot{\texttt{male}} &
\rot{\texttt{hhsize}} &
\rot{\texttt{treatment}}
   \\ \hline
\multirow{9}{*}{\begin{turn}{90}\texttt{health}\end{turn}} & NLMS$^*$ &  0.176 &  1.000 & -0.087 &  0.236 &  0.072 & -0.128 &  0.102 & -0.289 & -0.231 & -0.282 & -0.049 & -0.028 & --- \\
 & Control$^*$ & --- &  1.000 & -0.062 &  0.214 & -0.020 &  0.016 & -0.003 & -0.163 & -0.173 & -0.234 & -0.014 & -0.066 & \multirow{2}{*}{-0.049} \\
 & Treatment$^*$ & --- &  1.000 &  0.001 &  0.190 &  0.004 & -0.012 & -0.009 & -0.148 & -0.143 & -0.217 & -0.017 & -0.054 & \\
 & NLMS$^\dag$ &  0.207 &  1.000 &  0.052 &  0.346 & -0.015 & -0.049 &  0.093 & -0.221 & -0.129 & -0.304 & -0.036 & -0.105 & --- \\
 & Control$^\dag$ & --- &  1.000 & -0.062 &  0.221 & -0.023 &  0.012 &  0.002 & -0.164 & -0.157 & -0.227 & -0.002 & -0.079 & \multirow{2}{*}{-0.055} \\
 & Treatment$^\dag$ & --- &  1.000 & -0.005 &  0.206 & -0.002 & -0.012 & -0.001 & -0.154 & -0.131 & -0.213 &  0.003 & -0.064 & \\
 & NLMS$^\ddag$ &  0.176 &  1.000 & -0.087 &  0.236 &  0.072 & -0.128 &  0.102 & -0.289 & -0.231 & -0.282 & -0.049 & -0.028 & --- \\
 & Control$^\ddag$ & --- &  1.000 & -0.062 &  0.214 & -0.020 &  0.016 & -0.003 & -0.163 & -0.173 & -0.234 & -0.014 & -0.066 & \multirow{2}{*}{-0.049} \\
 & Treatment$^\ddag$ & --- &  1.000 &  0.001 &  0.190 &  0.004 & -0.012 & -0.009 & -0.148 & -0.143 & -0.217 & -0.017 & -0.054 & \\
\hline
\multirow{9}{*}{\begin{turn}{90}\texttt{insur\_any}\end{turn}} & NLMS$^*$ & -0.004 & -0.087 &  1.000 &  0.092 & -0.155 &  0.159 & -0.048 &  0.174 &  0.250 &  0.101 & -0.024 & -0.052 & --- \\
 & Control$^*$ & --- & -0.062 &  1.000 & -0.021 & -0.033 & -0.017 &  0.035 &  0.092 &  0.109 &  0.005 & -0.067 &  0.006 & \multirow{2}{*}{0.164} \\
 & Treatment$^*$ & --- &  0.001 &  1.000 &  0.053 & -0.095 &  0.060 &  0.015 &  0.060 & -0.080 & -0.102 & -0.044 & -0.045 & \\
 & NLMS$^\dag$ &  0.026 &  0.052 &  1.000 &  0.050 & -0.128 &  0.097 &  0.016 &  0.079 &  0.131 & -0.028 & -0.071 & -0.003 & --- \\
 & Control$^\dag$ & --- & -0.062 &  1.000 & -0.026 & -0.014 & -0.029 &  0.024 &  0.086 &  0.077 & -0.011 & -0.063 &  0.016 & \multirow{2}{*}{0.185} \\
 & Treatment$^\dag$ & --- & -0.005 &  1.000 &  0.049 & -0.091 &  0.057 &  0.019 &  0.041 & -0.118 & -0.102 & -0.030 & -0.044 & \\
 & NLMS$^\ddag$ &  0.008 & -0.087 &  1.000 &  0.140 & -0.222 &  0.212 & -0.036 &  0.239 &  0.281 &  0.091 & -0.039 & -0.100 & --- \\
 & Control$^\ddag$ & --- & -0.062 &  1.000 & -0.022 & -0.034 & -0.016 &  0.035 &  0.092 &  0.110 &  0.005 & -0.065 &  0.007 & \multirow{2}{*}{0.165} \\
 & Treatment$^\ddag$ & --- &  0.001 &  1.000 &  0.054 & -0.096 &  0.060 &  0.015 &  0.060 & -0.079 & -0.101 & -0.046 & -0.047 & \\
\hline
\end{tabular}
}
\caption{Pairwise correlations for the five intermediate variables with the variables that are observed in both datasets (i.e., the covariates and the intermediate variables) while calculated across various sets of study subjects. ``Control'' represents Oregon respondents who were not selected in the lottery, whereas ``Treatment'' represents Oregon respondents who were selected in the lottery. The column labeled ``\texttt{treatment}'' indicates the correlation between the treatment status (lottery winners = 1, non-winners = 0) and the respective intermediate variable for relevant OHIE cases.} \label{bigtable1a8}
\begin{tablenotes}
\item[*] Only respondents eligible for the ``Primary'' OHIE/NLMS data fusion model are used in calculation of these values.  The same set of respondents also defines those eligible for the ``Drop Health'' and ``No Covariates'' models. 
\item[\dag] Only respondents eligible for the ``Reduce NLMS'' OHIE/NLMS data fusion model (i.e., those at or below 200\% the poverty line) are used in calculation of these values.  
\item[\ddag] Only respondents eligible for the ``Drop Insurance'' OHIE/NLMS data fusion model (i.e., those without a missing value of \texttt{health}) are used in calculation of these values. 
\end{tablenotes}
\end{threeparttable}
\end{table}
\end{landscape}



\begin{center}
{\LARGE \bf Supplementary Materials: \\
Data Fusion for Predicting Long-Term Program Impacts
} \\
\vspace{.12in}
{\large Michael W.~Robbins, Sebastian Bauhoff, and Lane Burgette}
\end{center}


\renewcommand{\theequation}{A.\arabic{equation}}
\renewcommand{\thetable}{A.\arabic{table}}
\renewcommand{\thefigure}{A.\arabic{figure}}
\renewcommand{\thesection}{A.\arabic{section}}
\setcounter{equation}{0}
\setcounter{figure}{0}
\setcounter{table}{0}
\setcounter{section}{0}

\section{Additional Examples of Causal Pathways in the Context of Data Fusion} \label{causal}

Here, we expand upon the examples presented in Section \ref{examples}. First, we review the underlying notation and conditional relationships outlined therein.

First, assume that there is a single outcome denoted $Y$ and that there is a single intermediate (surrogate) variable $X$.  Lastly, denote the intervention by $Z$.  We will assume all variables can be related by linear conditional models.  We are primarily interested in
\begin{equation} \label{YgivenZ}
E[Y|Z] = \Phi_0 + \Phi_1 Z.
\end{equation}
Further, denote the conditional expectation of the surrogate given the intervention by
\begin{equation} \label{XgivenZ}
E[X|Z] = \phi_0 + \phi_1 Z.
\end{equation}
The quantity that cannot be estimated directly in the data fusion framework is
\begin{equation} \label{YgivenZX}
E[Y|Z,X] = \theta_0 + \theta_1 Z + \theta_2 X.
\end{equation}
In the event that (\ref{XgivenZ}) and (\ref{YgivenZX}) are known, it follows that
\begin{equation} \label{Phi1}
\Phi_1 = \theta_1 + \theta_2 \phi_1,
\end{equation}
which denotes the effect of the treatment on the outcome.  Further, $\Phi_0 = \theta_0 + \theta_2 \phi_0$.

Under the data fusion framework, (\ref{XgivenZ}) is estimated using the intervention dataset.  To circumvent the issue that (\ref{YgivenZX}) cannot be estimated directly, data fusion mandates that $\theta_1 = 0$ since $E|Y|Z,X] = \theta_0 + \theta_2 X$ can be estimated using an outcomes dataset.  When $\theta_1 = 0$, the effect of the treatment on the outcome is given by $\Phi_1 = \theta_2 \phi_1$ (where $\theta_2$ is estimated using the outcomes dataset and $\phi_1$ is estimated using the intervention dataset).  The data fusion ``estimator'' of $\Phi_1$ is given by
\begin{equation} \label{tildePhi1}
\widetilde{\Phi}_1 = \theta_2 \phi_1.
\end{equation}
The scaled bias in the data fusion estimator is given by
\begin{equation} \label{bias}
\mbox{bias} = \frac{\widetilde{\Phi}_1-\Phi_1}{\Phi_1} = -\frac{\theta_1}{\theta_1 + \theta_2 \phi_1}.
\end{equation}
Additionally, the portion of the treatment effect explained by the surrogate is
\begin{equation} \label{portion}
\mbox{portion} = \frac{\Phi_1 - \theta_1}{\Phi_1} = 1+\mbox{bias}.
\end{equation}

\noindent {\bf Example 3 (Mendelian Randomization):} \\

\begin{figure}[h]
\begin{center}
\begin{tikzpicture}
\matrix[row sep = 0.35cm, column sep = 0.35cm]
{
 \node[circle,thick, draw, white, line width=0.05cm] (n1) {\color{black} $Z$}; & & & & & & \node[circle,thick, draw, white, line width=0.05cm] (n2) {\color{black} $X$}; & & & & & &  \node[circle,thick, draw, white, line width=0.05cm] (n3) {\color{black} $Y$}; \\
 & & & & & &  & & & & & &   \\
 & & & & & &  & & & & & &   \\
 & & & & & &  & & & & & &   \\
& & & & & & \node[circle,thick, draw, white, line width=0.05cm] (n4) {\color{black} $U$}; & & & & & &  \\
};
\draw [->,black, thick, line width=0.05cm] (n1) -- node [above]    {}  (n2);
\draw [->,black, thick, line width=0.05cm] (n4) -- node [above]    {}  (n2);
\draw [->,black, thick, line width=0.05cm] (n4) -- node [above]    {}  (n3);
\end{tikzpicture}
\caption{Surrogate relates to the outcome only through an unobserved confounder, and the treatment affects only the surrogate.} \label{fig1}
\end{center}
\end{figure}
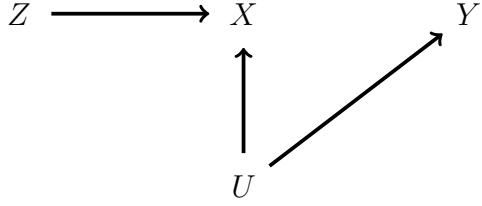

We consider settings that are consistent with the surrogate paradox and consider whether or not these settings violate the assumptions for data fusion.  First, we assume that $X$ and $Y$ are not causally related and that a correlation exists between the two of them due only to an unobserved confounder $U$ that ``causes'' both $X$ and $Y$.  Further, assume that the treatment $Z$ affects only the surrogate $X$ and has no causal relationship between $U$ or $Y$.  This setting represents a causal pathway that Mendelian randomization is designed to address and is a simplification of Example 2.  The relationship between the variables $Z$, $X$, $U$, and $Y$ is illustrated in Figure \ref{fig1}.

The stochastic mechanism that generates the variables $U$, $X$, and $Y$ given $Z$ in this example is expressed using the following conditional distributions:
\begin{eqnarray*}
f(U|Z):& &U = \alpha_0 + \epsilon_u, \\
f(X|Z,U): & &X= \beta_0 + \beta_1 U + \beta_2 Z + \epsilon_x, \\
f(Y|Z,U,X): & & Y =\gamma_0 + \gamma_1 U + \epsilon_y,
\end{eqnarray*}
where $\epsilon_u$, $\epsilon_x$, and $\epsilon_y$ are normal random errors with mean zero and variances $\sigma^2_u$, $\sigma^2_x$, and $\sigma^2_y$, respectively.

The conditional distribution in (\ref{XgivenZ}) has $\phi_1 = \beta_2$.  Further, it can be shown that the conditional distribution in (\ref{YgivenZX}) is expressed by
\[
E[Y|Z,X] = c_0 - \beta_2 c_1 Z + c_1 X.
\]
where
\[
c_1 = \frac{\gamma_1 \beta_1 \sigma^2_u}{\beta_1^2\sigma_u^2+\sigma_x^2}
~~~\mbox{and}~~~
c_0 = \gamma_0+\gamma_1 \alpha_0 +c_1(\beta_0 + \beta_2 \alpha_0).
\]
That is, $\theta_1 = -\beta_2c_1$ and $\theta_2 = c_1$.  Note that $\theta_1 \neq 0$, which implies that the stated assumptions for data fusion are violated.  Using (\ref{Phi1}), we see that
\[
\Phi_1 = 0,
\]
and from (\ref{tildePhi1}) it follows that
\[
\widetilde{\Phi}_1=\beta_2c_1.
\]
Both the bias in (\ref{bias}) and the portion in (\ref{portion}) are infinite since $\Phi_1=0$.  \\

\noindent {\bf Example 4 (Reverse causality):} \\

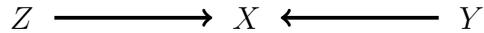
\begin{figure}[h]
\begin{center}
\begin{tikzpicture}
\matrix[row sep = 0.35cm, column sep = 0.35cm]
{
 \node[circle,thick, draw, white, line width=0.05cm] (n1) {\color{black} $Z$}; & & & & & & \node[circle,thick, draw, white, line width=0.05cm] (n2) {\color{black} $X$}; & & & & & &  \node[circle,thick, draw, white, line width=0.05cm] (n3) {\color{black} $Y$}; \\
};
\draw [->,black, thick, line width=0.05cm] (n1) -- node [above]    {}  (n2);
\draw [->,black, thick, line width=0.05cm] (n3) -- node [above]    {}  (n2);
\end{tikzpicture}
\caption{Reverse causality (the outcome has a causal effect on the surrogate).} \label{fig3}
\end{center}
\end{figure}

We now consider circumstances where the treatment has a causal effect on the surrogate and the outcome also has a causal effect on the surrogate.  However, the treatment does not directly affect the outcome, nor does the surrogate have a causal effect on the outcome.  This causal pathyway embodies ``reverse causality'' and is illustrated in Figure \ref{fig3}.

The stochastic mechanism that generates the variables $X$ and $Y$ given $Z$ in this example is expressed using the following conditional distributions:
\begin{eqnarray*}
f(X|Z,Y): & &X= \beta_0 +  \beta_2 Z + \beta_3Y + \epsilon_x, \\
f(Y|Z): & & Y =\gamma_0 + \epsilon_y,
\end{eqnarray*}
where $\epsilon_x$ and $\epsilon_y$ are independent error terms with variances $\sigma^2_x$ and $\sigma^2_y$, respectively.
It follows that $\phi_1 = \beta_2$ with
\[
\theta_1 = - c_2 \beta_2
~~~~\mbox{and}~~~~
\theta_2 = c_2
\]
where
\[
c_2=\frac{\beta_3 \sigma^2_y}{\beta_3^2\sigma_y^2+\sigma_x^2}.
\]
Again, we see $\theta_1\neq 0$, implying the conditions for data fusion are not satisfied.
The effect of the treatment on the outcome is given by
\[
\Phi_1 = 0
~~~~\mbox{and}~~~~
\widetilde{\Phi}_1 = c_2 \beta_2.
\]
As in Example 3, the scaled bias is infinite since $\Phi_1=0$. \\

\noindent {\bf Example 5:} \\

\begin{figure}[h]
\begin{center}
\begin{tikzpicture}
\matrix[row sep = 0.35cm, column sep = 0.35cm]
{
 \node[circle,thick, draw, white, line width=0.05cm] (n1) {\color{black} $Z$}; & & & & & & \node[circle,thick, draw, white, line width=0.05cm] (n2) {\color{black} $X$}; & & & & & &  \node[circle,thick, draw, white, line width=0.05cm] (n3) {\color{black} $Y$}; \\
 & & & & & &  & & & & & &   \\
 & & & & & &  & & & & & &   \\
 & & & & & &  & & & & & &   \\
& & & & & & \node[circle,thick, draw, white, line width=0.05cm] (n4) {\color{black} $U$}; & & & & & &  \\
};
\draw [->,black, thick, line width=0.05cm] (n1) -- node [above]    {}  (n2);
\draw [->,black, thick, line width=0.05cm] (n4) -- node [above]    {}  (n2);
\draw [->,black, thick, line width=0.05cm] (n4) -- node [above]    {}  (n3);
\draw [->,black, thick, line width=0.05cm] (n2) -- node [above]    {}  (n3);
\draw [->,black, thick, line width=0.05cm] (n1) edge [bend left] node [above]    {}  (n3);
\end{tikzpicture}
\caption{Surrogate relates to the outcome directly and through an unobserved confounder, and the treatment affects the surrogate and the outcome directly.} \label{fig5}
\end{center}
\end{figure}
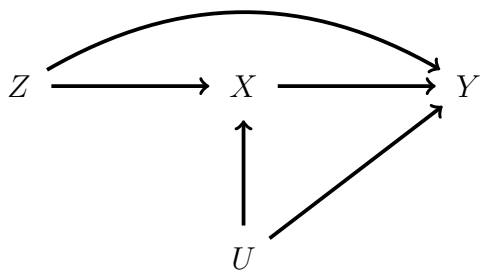

We now consider a scenario that is a combination of Examples 1 and 2.  Specifically, this example is the same as Example 3, but we now assume that the intervention directly affects the outcome through a pathway that is not explained by the surrogate or by the unobserved confounder.  Graphically, the causal relationship considered here is shown in Figure \ref{fig5}.

The stochastic mechanism that generates the variables $U$, $X$, and $Y$ given $Z$ in this example is expressed using the following conditional distributions:
\begin{eqnarray*}
f(U|Z):& &U = \alpha_0 + \epsilon_u, \\
f(X|Z,U): & &X= \beta_0 + \beta_1 U + \beta_2 Z + \epsilon_x, \\
f(Y|Z,U,X): & & Y =\gamma_0 + \gamma_1 U + \gamma_2 X + \gamma_3Z+ \epsilon_y,
\end{eqnarray*}

The conditional model in (\ref{XgivenZ}) has $\phi_1 = \beta_2$.  Likewise, for the model in (\ref{YgivenZX}), we see
\[
\theta_1 = \gamma_3-c_1 \beta_2
~~~~\mbox{and}~~~~
\theta_2 = c_1+\gamma_2,
\]
where
\[
c_1=\frac{\gamma_1 \beta_1 \sigma^2_u}{\beta_1^2\sigma_u^2+\sigma_x^2}.
\]
Since $\theta_1 \neq 0$, the conditions for data fusion are not satisfied.
The treatment effect in (\ref{YgivenZ}) is given by
\[
\Phi_1 = \gamma_3+ \beta_2\gamma_2
~~~~\mbox{with}~~~~
\widetilde{\Phi}_1 = (c_1+\gamma_2)\beta_2.
\]
Therefore, the bias in (\ref{bias}) is represented by
\[
\mbox{bias}=\frac{\beta_2c_1-\gamma_3}{\gamma_3+\gamma_2\beta_2}.
\]
Note that if $\beta_2c_1-\gamma_3 =0$, the bias is zero (and therefore the portion is 1).  This is possible even if all relationships (i.e., coefficients) are positive.  Similarly, if $\beta_2c_1-\gamma_3 =0$ and $\gamma_2=0$, the surrogate none of the treatment effect, yet the portion in (\ref{portion}) is calculated as being one.  Note also that it is possible to have $\Phi_1 < 0$ if all correlations between variables are positive other than the latent pathway between the outcome and intervention (which is controlled by $\gamma_3$)---This is the so-called surrogate paradox.\\

\noindent {\bf Example 6 (Multivariate Surrogate):} \\

\begin{figure}[h]
\begin{center}
\begin{tikzpicture}
\matrix[row sep = 0.35cm, column sep = 0.35cm]
{
 \node[circle,thick, draw, white, line width=0.05cm] (n1) {\color{black} $Z$}; & & & & & & \node[circle,thick, draw, white, line width=0.05cm] (n2) {\color{black} $X_2$}; & & & & & &  \node[circle,thick, draw, white, line width=0.05cm] (n3) {\color{black} $Y$}; \\
 & & & & & &  & & & & & &   \\
 & & & & & &  & & & & & &   \\
 & & & & & &  & & & & & &   \\
& & & & & & \node[circle,thick, draw, white, line width=0.05cm] (n4) {\color{black} $X_1$}; & & & & & &  \\
};
\draw [->,black, thick, line width=0.05cm] (n1) -- node [above]    {}  (n2);
\draw [->,black, thick, line width=0.05cm] (n4) -- node [above]    {}  (n2);
\draw [->,black, thick, line width=0.05cm] (n4) -- node [above]    {}  (n3);
\draw [->,black, thick, line width=0.05cm] (n2) -- node [above]    {}  (n3);
\end{tikzpicture}
\caption{The case of two observed intermediate variables, where only the second is influenced by the intervention.} \label{fig4}
\end{center}
\end{figure}
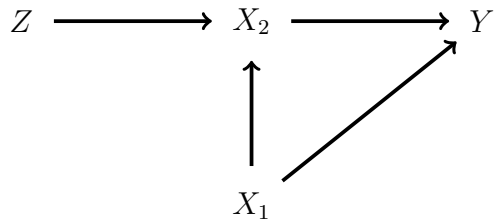

The next example is the same as Example 2, except we now assume that the the unobserved confounder is observed.  This essentially implies that we have two intermediate variables denoted ${\bf X}=(X_1,X_2)'$, where $X_1$ is equivalent to $U$ in Example 2 and $X_2$ is equivalent to $X$.  Graphically, the causal relationship considered here is shown in Figure \ref{fig4}.

The stochastic mechanism that generates the variables $X_1$, $X_2$, and $Y$ given $Z$ in this example is expressed using the following conditional distributions:
\begin{eqnarray*}
f(X_1|Z):& &X_1 = \alpha_0 + \epsilon_1, \\
f(X_2|Z,X_1): & &X_2= \beta_0 + \beta_1 X_1 + \beta_2 Z + \epsilon_2, \\
f(Y|Z,X_1,X_2): & & Y =\gamma_0 + \gamma_1 X_1 + \gamma_2 X_2 + \epsilon_y,
\end{eqnarray*}
where $\epsilon_1$ and $\epsilon_2$ are independent Gaussian errors with respective variances $\sigma_1^2$ and $\sigma_2^2$.  Substitution shows that
\[
E[Y|Z]= \gamma_0+\gamma_1 \alpha_0 + \gamma_2 \beta_0 + \gamma_2 \beta_1 \alpha_0 + \gamma_2 \beta_2 Z,
\]
which implies that
\[
\Phi_1 = \gamma_2\beta_2,
\]
as seen in Example 2.

This example is handled in the data fusion context as follows. First, we write the joint distribution of ${\bf X}$ given $Z$:
\[
\left[  \begin{array}{c} X_1 \\ X_2 \end{array}\right]  = \left[  \begin{array}{c} \alpha_0 \\ \beta_0 + \beta_1 \alpha_0 \end{array}\right] + \left[  \begin{array}{c} 0 \\ \beta_2 \end{array}\right]Z + \bm{\epsilon}
~~~~
\Rightarrow
~~~~
\bm{X} = \bm{\beta}_0^T + \bm{\beta}^T Z + \bm{\epsilon}
\]
where $\bm{\beta} = (0,\beta_2)$ and
\[
\bm{\epsilon} \sim \mbox{N}\left(\bm{0}, \left[  \begin{array}{cc} \sigma_1^2 & \beta_1 \sigma_2^2 \\  \beta_1 \sigma_2^2  &  \beta_1^2 \sigma_1^2 +\sigma_2^2 \end{array}\right] \right).
\]
Further, we write $f(Y|Z,X_1,X_2)$ as $Y=\gamma_0 + \bm{\gamma}^T{\bf X} + \epsilon_y$ where $\bm{\gamma}=(\gamma_1,\gamma_2)'$.  In the data fusion framework, $\bm{\gamma}$ is estimated using the outcomes dataset and $\bm{\beta}$ is estimated using the intervention dataset.  The data fusion estimate of $\Phi_1$ is
\[
\widetilde{\Phi}_1 = \bm{\beta}\bm{\gamma} = \left[  \begin{array}{cc} 0 & \beta_2 \end{array}\right]\left[  \begin{array}{c} \gamma_1 \\ \gamma_2 \end{array}\right] = \beta_2 \gamma_2,
\]
which exactly equals the true value of $\Phi_1$.

\end{document}